\documentclass[manuscript]{aastex}

\shorttitle{Supernovae and Protoplanetary Disks}
\shortauthors{Ouellette, Desch & Hester}

\newcommand{\msol}{\mbox{${\rm M}_{\odot}$}} 
\newcommand{\hii}{\mbox{${\rm H}\,${\sc ii}}}
\newcommand{\vz}{\mbox{$\langle v_{z} \rangle$}} 
\newcommand{\fesixty}{\mbox{${}^{60}{\rm Fe}$}} 
\newcommand{\feratio}{\mbox{${}^{60}{\rm Fe}/{}^{56}{\rm Fe}$}} 
\newcommand{\gtsimeq}{\raisebox{-0.6ex}{$\, \stackrel{\raisebox{-.2ex}%
{$\textstyle >$}}{\sim}\,$}}
\newcommand{\ltsimeq}{\raisebox{-0.6ex}{$\, \stackrel{\raisebox{-.2ex}%
{$\textstyle <$}}{\sim}\,$}}

\begin{document}


\title{Interaction of Supernova Ejecta with Nearby Protoplanetary Disks}


\author{N. Ouellette}
\affil{Department of Physics, Arizona State University, PO Box 871504, Tempe, AZ 85287-1504}
\author{S. J. Desch and J. J. Hester}
\affil{School of Earth and Space exploration, Arizona State University, PO Box 871404, Tempe, AZ 85287-1404}



\begin{abstract}

The early Solar System contained short-lived radionuclides such as ${}^{60}{\rm Fe}$
($t_{1/2} = 1.5$ Myr) whose most likely source was a nearby supernova.
Previous models of Solar System formation considered a supernova shock that triggered 
the collapse of the Sun's nascent molecular cloud.
We advocate an alternative hypothesis, that the Solar System's protoplanetary disk
had already formed when a very close ($< 1$ pc) supernova injected radioactive material 
directly into the disk.
We conduct the first numerical simulations designed to answer two questions related 
to this hypothesis: will the disk be destroyed by such a close supernova; and will 
any of the ejecta be mixed into the disk?
Our simulations demonstrate that the disk does not absorb enough momentum from the 
shock to escape the protostar to which it is bound.
Only low amounts ($< 1\%$) of mass loss occur, due to stripping by Kelvin-Helmholtz 
instabilities across the top of the disk, which also mix into the disk about 1\% of 
the intercepted ejecta.
These low efficiencies of destruction and injectation are due to the fact that the 
high disk pressures prevent the ejecta from penetrating far into the disk before stalling.
Injection of gas-phase ejecta is too inefficient to be consistent with the abundances of 
radionuclides inferred from meteorites.
On the other hand, the radionuclides found in meteorites would have condensed into dust 
grains in the supernova ejecta, and we argue that such grains will be injected directly 
into the disk with nearly 100\% efficiency. 
The meteoritic abundances of the short-lived radionuclides such as ${}^{60}{\rm Fe}$ therefore
are consistent with injection of grains condensed from the ejecta of a nearby ($< 1$ pc)
supernova, into an already-formed protoplanetary disk.
\end{abstract}


\section{Introduction}

Many aspects of the formation of the Solar System are fundamentally affected by
the Sun's stellar birth environment, but to this day the type of environment has 
not been well constrained.  
Did the Sun form in a quiescent molecular cloud like the Taurus molecular cloud
in which many T Tauri stars are observed today?
Or did the Sun form in the vicinity of massive O stars that ionized surrounding gas,
creating an $\hii$ region before exploding as core-collapse supernovae? 
Recent isotopic analyses of meteorites reveal that the early Solar System held live 
$\fesixty$ at moderately high abundances, ${}^{60}{\rm Fe} / {}^{56}{\rm Fe} \sim 
3 - 7 \times 10^{-7}$ (Tachibana \& Huss 2003; Huss \& Tachibana 2004; Mostefaoui 
et al.\ 2004, 2005; Quitte et al.\ 2005; Tachibana et al.\ 2006).  
Given these high initial abundances, the origin of this short-lived radionuclide (SLR), 
with a half-life of 1.5 Myr, is almost certainly a nearby supernova, and these 
meteoritic isotopic measurements severely constrain the Sun's birth environment.

Since its discovery, the high initial abundance of $\fesixty$ in the early Solar System 
has been recognized as demanding an origin in a nearby stellar nucleosynthetic source, 
almost certainly a supernova (Jacobsen 2005; Goswami et al.\ 2005; Ouellette et al.\ 2005; 
Tachibana et al.\ 2006, Looney et al.\  2006).
Inheritance from the interstellar medium (ISM) can be ruled out: the average abundance
of $\fesixty$ maintained by ongoing Galactic nucleosynthesis in supernovae and asymptotic-giant-branch 
(AGB) stars is estimated at $\feratio = 3 \times 10^{-8}$ (Wasserburg et al.\ 1998) to 
$3 \times10^{-7}$ (Harper 1996), lower than the meteoritic ratio.
Moreover, this $\fesixty$ is injected into the hot phase of the ISM (Meyer \& Clayton 2000),
and incorporation into molecular clouds and solar systems takes $\sim 10^{7}$ years or more
(Meyer \& Clayton 2000; Jacobsen 2005), by which time the $\fesixty$ has decayed.
A late source is argued for (Jacobsen 2005; see also Harper 1996, Meyer \& Clayton 2000).
Production within the Solar System itself by irradiation of rocky material by solar energetic 
particles has been proposed for the origin of other SLRs (e.g., Lee et al.\ 1998; Gounelle et al.\ 2001),
but neutron-rich $\fesixty$ is produced in very low yields by this process.
Predicted abundances are $\feratio \sim 10^{-11}$, too low by orders of magnitude to explain the
meteoritic abundance (Lee et al.\ 1998; Leya et al.\ 2003; Gounelle et al.\ 2006).
The late source is therefore a stellar nucleosynthetic source, either a supernova or an AGB 
star.  
AGB stars are not associated with star-forming regions: 
Kastner \& Myers (1994) used astronomical observations to estimate a firm upper limit of $\approx 3 \times 10^{-6}$
per Myr to the probability that our Solar System was contaminated by material from an AGB star.
The yields of $\fesixty$ from an AGB star also may not be sufficient to explain the meteoritic ratio
(Tachibana et al.\ 2006). 
Supernovae, on the other hand, are commonly associated with star-forming regions, and a 
core-collapse supernova is by far the most plausible source of the Solar System's $\fesixty$.

Supernovae are naturally associated with star-forming regions because the typical lifetimes
of the stars massive enough to explode as supernovae ($\gtsimeq 8 \, M_{\odot}$) are $\ltsimeq 10^{7}$ yr,
too short a time for them to disperse away from the star-forming region they were born in.
Low-mass ($\sim 1 \, M_{\odot}$) stars are also born in such regions.
In fact, astronomical observations indicate that the {\it majority} of low-mass stars form in association
with massive stars. 
Lada \& Lada (2003) conducted a census of protostars in deeply embedded clusters complete to 2 kpc
and found that 70-90\% of stars form in clusters with $>$ 100 stars.
Integration of the cluster initial mass function indicates that of all stars born in clusters of at
least 100 members, about 70\% will form in clusters with at least one star massive enough to supernova
(Adams \& Laughlin 2001; Hester \& Desch 2005).
Thus at least 50\% of all low-mass stars form in association with a supernova, and it is reasonable to
assume the Sun was one such star.  
Astronomical observations are consistent with, and the presence of $\fesixty$ demands, formation of the 
Sun in association with at least one massive star that went supernova.

While the case for a supernova is strong, constraining the proximity and the timing of the supernova 
is more difficult. 
The SLRs in meteorites provide some constraints on the timing.
The SLR $\fesixty$ must have made its way from the supernova to the Solar System in only a few half-lives;
models in which multiple SLRs are injected by a single supernova provide a good match to meteoritic
data only if the meteoritic components containing the SLRs formed $\ltsimeq 1$ Myr after the supernova
(e.g., Meyer 2005, Looney et al. 2007). 
The significance of this tight timing constraint is that the formation of the Solar System was somehow 
associated with the supernova.
Cameron \& Truran (1977) suggested that the formation of the Solar System was triggered by the shock 
wave from the same supernova that injected the SLRs, and subseqeuent numerical simulations show this is
a viable mechanism, provided several parsecs of molecular gas lies between the supernova and the Solar System's
cloud core, or else the supernova shock will shred the molecular cloud (Vanhala \& Boss 2000, 2002).
The likelihood of this initial condition has not yet been established by astronomical observations. 
Also in 1977, T.\ Gold proposed that the Solar System acquired its radionuclides from a nearby supernova,
{\it after} its protoplanetary disk had already formed (Clayton 1977).
Astronomical observations strongly support this scenario, especially since protoplanetary disks were 
directly imaged $\sim 0.2$ pc from the massive star $\theta^{1}$ Ori C in the Orion Nebula (McCaughrean
\& O'Dell 1996).
Further imaging has revealed protostars with disks near ($\le 1\, {\rm pc}$) massive stars in the Carina Nebula (Smith et al.\ 
2003), NGC 6611 (Oliveira et al.\ 2005), and M17 and Pismis 24 (de Marco et al.\ 2006). 
This hypothesis, that the Solar System acquired SLRs from a supernova that occurred $< 1$ pc away, after
the Sun's protoplanetary disk had formed, is the focus of this paper. 

In this paper we address two main questions pertinent to this model.
First, are protoplanetary disks destroyed by the explosion of a supernova a fraction of a parsec away?
Second, can supernova ejecta containing SLRs be mixed into the disk? 
These questions were analytically examined in a limited manner by Chevalier (2000).
Here we present the first multidimensional numerical simulations of the interaction of supernova ejecta with 
protoplanetary disks. 
In \S 2 we describe the numerical code, Perseus, we have written to study this problem.
In \S 3 we discuss the results of one canonical case in particular, run at moderate spatial resolution.
We examine closely the effects of our limited numerical resolution in \S 4, and show that we have achieved
sufficient convergence to draw conclusions about the survivability of protoplanetary disks hit by supernova
shocks. 
We conduct a parameter study, investigating the effects of supernova energy and distance and disk mass, as 
described in \S 5.
Finally, we summarize our results in \S 6, in which we conclude that disks are not destroyed by a nearby supernova,
that gaseous ejecta is not effectively mixed into the disks, but that solid grains from the supernova likely are,
thereby explaining the presence of SLRs like $\fesixty$ in the early Solar System. 

\section{Perseus}

We have written a 2-D (cylindrical) hydrodynamics code we call Perseus.  
Perseus (son of Zeus) is based heavily on the Zeus algorithms (Stone \& Norman 1992).  
The code evolves the system while obeying the equations of conservation of  mass, momentum and energy:
\begin{equation}
\frac{D\rho}{Dt}+\rho\nabla\cdot \vec{v}=0
\end{equation}
\begin{equation}
\rho\frac{Dv}{Dt}=-\nabla p-\rho\nabla\Phi
\end{equation}
\begin{equation}
\rho\frac{D}{Dt}\left(\frac{e}{\rho}\right)=-p\nabla\cdot \vec{v},
\end{equation}
where $\rho$ is the mass density, $\vec{v}$ is the velocity, $p$ is the pressure, $e$ is the internal 
energy density and $\Phi$ is the gravitational potential (externally imposed).  
The Lagrangean, or comoving derivative $D/Dt$ is defined as
\begin{equation}
\frac{D}{Dt}\equiv \frac{\partial}{\partial t} + \vec{v}\cdot\nabla.
\end{equation}
The pressure and energy are related by the simple equation of state appropriate for the ideal gas law, 
$p=e(\gamma -1)$, where $\gamma$ is the adiabatic index.  
The term $p\nabla\cdot \vec{v}$ represents mechanical work.  

Currently, the only gravitational potential $\Phi$ used is a simple point source, representing a star 
at the center of a disk.
This point mass is constrained to remain at the origin.  
Technically this violates conservation of momentum by a minute amount by excluding the gravitational
force of the disk on the central star.
As discussed in \S 4, the star should acquire a velocity $\sim 10^{2} \, {\rm cm} \, {\rm s}^{-1}$ at
the end of our simulations.
In future simulations we will include this effect, but for the problem explored here this is completely
negligible. 

The variables evolved by Perseus are set on a cylindrical grid.  
The  program is separated in two steps: the source and the transport step.  
The source step calculates the changes in velocity and energy due to sources and sinks.  
Using finite difference approximations, it evolves $\vec{v}$ and $e$ according to
\begin{equation}
\rho \frac{\partial \vec{v}}{\partial t}=-\nabla p - \rho\nabla\Phi -\nabla \cdot \mbox{\boldmath$Q$}
\end{equation}
and
\begin{equation}
\rho \frac{\partial e}{\partial t}=-p\nabla\cdot\vec{v} - \mbox{\boldmath$Q$}: \nabla\vec{v},
\end{equation}
where \mbox{\boldmath$Q$} is the {\it tensor} artificial viscosity.  
Detailed expressions for the artificial viscosity can be found in Stone \&  Norman (1992).

The transport step evolves the variables according to the velocities present on the grid.  
For a given variable $A$, the conservation equation is solved, using finite difference approximations:
\begin{equation}
\frac{d}{dt} \int_{V}A\, dV=-\oint_{S}A\, \vec{v}\cdot d\vec{S}.
\end{equation}
The variables $A$ advected in this way are density $\rho$, linear and angular momentum $\rho\vec{v}$ 
and $R \rho v_{\phi}$, and energy density $e$.  
As in the Zeus code, 
$A$ on each surface element is found with an upwind interpolation scheme; we use 
second-order van Leer interpolation.

Perseus is an explicit code and must satisfy the Courant-Friedrichs-Lewis (CFL) stability criterion.
The amount of time advanced per timestep, essentially,  must not exceed the time it could take 
for information to cross a grid zone in the physical system.
In every grid zone, the thermal time step $\delta t_{c_s}=\Delta x/(c_s)$ is computed, where 
$\Delta x$ is the size of the zone (smallest of the $r$ and $z$ dimension) and $c_s$ is the sound speed.
Also computed are $\delta t_{\rm r}=\delta r/(|v_{\rm r}|)$ and $\delta t_{\rm z}=\delta z/(|v_{\rm z}|)$, where 
$\Delta r$ and $\Delta z$ are the sizes of the zone in the $r$ and $z$ directions respectively.  
Because of artificial viscosity, a viscous time step must also be added for stability.   
For a given grid zone, the viscous time step $\delta t_{\rm visc}={\rm max}(|(l\; \nabla\cdot\vec{v}/\delta r^2)|$, 
$|(l\; \nabla\cdot\vec{v}/\delta z^2)|$) is computed, where $l$ is a length chosen to be a 3 zone widths.  
The final $\Delta t$ is taken to be 
\begin{equation}
\Delta t=C_0 \, (\delta t_{c_s}^{-2}+\delta t_{\rm r}^{-2}+\delta t_{\rm z}^{-2}+\delta t_{\rm visc}^{-2})^{-1/2},
\end{equation}
where $C_0$ is the Courant number, a safety factor, taken to be $C_0$=0.5.  
To insure stability, $\Delta t$ is computed over all zones, and the smallest value is kept for the next step of the simulation.

Boundary conditions were implemented using ghost zones as in the Zeus code. 
To allow for supernova ejecta to flow past the disk, inflow boundary conditions were used at the upper boundary 
($z=z{\rm _{max}}$), and outflow boundary conditions were used at the lower boundary ($z=z{\rm _{min}}$) and outer boundary 
($r=r{\rm _{max}}$).
Reflecting boundary conditions, were used on the inner boundary ($r=r{\rm _{min}} \neq 0$) to best model the 
symmetry about the protoplanetary disk's axis. 
The density and velocity of gas flowing into the upper boundary were varied with time to match the ejecta properties 
(see \S 3). 

A more detailed description of the algorithms used in Perseus can be found in Stone \&  Norman (1992).

\subsection{Additions to Zeus}

To consider the particular problem of high-velocity ejecta hitting a protoplanetary disk, we wrote Perseus with the 
following additions to the Zeus code.
One minor change is the use of a non-uniform grid.
In all of our simulations we used an orthogonal grid with uniform spacing in $r$ but non-uniform spacing in the
$z$ direction.
For example, in the canonical simulation (\S 3), the computational domain extends from $r = 4$ to 80 AU, with
spacing $\Delta r = 1 \, {\rm AU}$, for a total of 76 zones in $r$.
The computational domain extends from $z = -50 \, {\rm AU}$ to $+90 \, {\rm AU}$, but zone spacings vary with $z$,
from $\Delta z = 0.2 \, {\rm AU}$ at $z = 0$, to $\Delta z \approx 3 \, {\rm AU}$ at the upper boundary.
Grid spacings increased geometrically by 5\% per zone, for a total of 120 zones in $z$. 

Another addition was the use of a radiative cooling term.
The simulations bear out the expectation that almost all of the shocked supernova ejecta flow past the disk
before they have time to cool significantly. 
Cooling is significant only where the ejecta collide with the dense gas of the disk itself, but there the 
cooling is sensitive to many unconstrained physical properties to do with the the chemical state of the gas,
properties of dust, etc. 
To capture the gross effects of cooling (especially compression of gas near the dense disk gas) in a computationally
simple way, we have adopted the following additional term in the energy equation, implemented in the source step:
\begin{equation}
\frac{\partial e}{\partial t} = -n_{\rm e} n_{\rm p} \Lambda,
\end{equation}
where $n{\rm_e}$ and $n{\rm_p}$ are the number of protons and electrons in the gas, and $\Lambda$ is the cooling function.  
The densities $n{\rm_e}$ and $n{\rm_p}$ are obtained simply by assuming the hydrogen gas is fully ionized, so 
$n{\rm_e} = n{\rm_p} = \rho / 1.4 m_{\rm H}$.  
For gas temperatures above $10^{4} \, {\rm K}$, we take $\Lambda$ of a solar-metallicity gas from Sutherland and Dopita
(1993); $\Lambda$ typically ranges between $10^{-24} \, {\rm erg} \, {\rm cm}^3 \, {\rm s}^{-1}$ (at $T = 10^4 \,{\rm K}$)
and $\Lambda = 10^{-21} \, {\rm erg} \, {\rm cm}^{3} \, {\rm s}^{-1}$ (at $T = 10^5 \, {\rm K}$).
Below $10^{4} \, {\rm K}$ we adopted a flat cooling function of $\Lambda = 10^{-24} \, {\rm erg} \, {\rm cm}^{3} \, {\rm s}^{-1}$.
At very low temperatures it is necessary to include heating processes as well as cooling, or else the gas rapidly cools to 
unreasonable temperatures. 
Rather than handle transfer of radiation from the central star, we defined a minimum temperature below which the gas is
not allowed to cool: $T_{\rm min} = 300 \, (r / 1 \, {\rm AU})^{-3/4} \, {\rm K}$. 
Perseus uses a simple first-order, finite-difference equation to handle cooling.  
Although this method is not as precise as a predictor-corrector method, in \S 2.4 we show that it is sufficiently accurate 
for our purposes. 

Because Perseus is an explicit code, the implementation of a cooling term demands the introduction of a cooling
time step to insure that the gas doesn't cool too rapidly during one time step, resulting in negative temperatures or 
other instabilities. 
For a radiating gas, the cooling timescale can be approximated by $t_{\rm cool} \approx k_B T / n \Lambda$, where 
$k_B$ is the Boltzmann constant, $T$ is the temperature of the gas, $n$ is the number density and $\Lambda$ is the 
appropriate cooling function.  
This cooling timescale is calculated on all the grid zones where the temperature exceeds $10^3\,{\rm K}$, and the
cooling time step $\delta t_{\rm cool}$ is defined to be 0.025 times the shortest cooling timescale on the grid.
If the smallest cooling time step is shorter that the previously calculated $\Delta t$ as defined by eq.\ [8], then
it becomes the new time step.  
We ignore zones where the temperature is below $10^{3} \, {\rm K}$ because heating and cooling are not fully 
calculated anyway, and because these zones are always associated with very high densities and cool extremely rapidly,
on timescales as short as hours, too rapidly to reasonably follow anyway.

Finally, to follow the evolution of the ejecta gas with respect to the disk gas, a tracer ``color density" was added.  
By defining a different density, the color density $\rho_{\rm c}$, it is possible to follow the mixing of a two specific 
parts of a system, in this case the ejecta and the disk.  
By comparing  $\rho_{\rm c}$ to  $\rho$, it is possible to know how much of the ejecta is present in a given zone 
relative to the original material.  
It is important to note that $\rho_{\rm c}$ is a tracer and does not affect the simulation in any way.

\subsection{Sod Shock-Tube}

\begin{figure}
\epsscale{.55}
\plotone{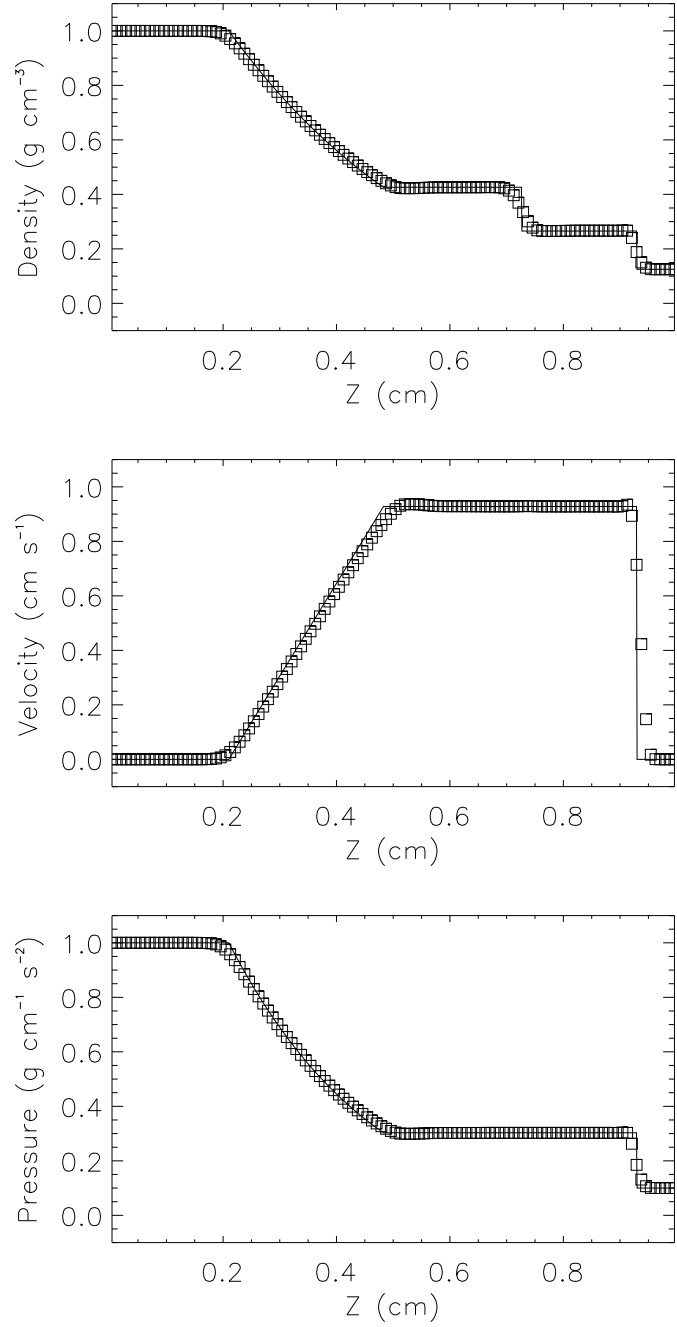}
\caption{Sod shock-tube problem benchmark.  The squares are the results of the simulation using Perseus, and the 
solid line is the analytical solution from Hawley et al.\ (1984).}
\label{Fig:1}
\end{figure}

We have benchmarked the Perseus code against a well-known analytic solution, the Sod shock tube (Sod 1978).  
Tests were performed to verify the validity of Perseus's results.  
It is a 1-D test, and hence was only done in the $z$ direction, as curvature effects would render this test 
invalid in the $r$ direction.   
Therefore, the gas was initially set spatially uniform in $r$. 
120 zones were used in the $z$ direction.
The other initial conditions of the Sod shock-tube are as follows:  the simulation domain is split in half and filled 
with a $\gamma$=1.4 gas;  in one half ($z < 0.5\,{\rm cm}$), the gas has a pressure of $1.0 \, {\rm dyne} \, {\rm cm}^{-2}$ 
and a density of $1.0\, {\rm g} \, {\rm cm}^{-3}$, while in the other half ($z > 0.5\,{\rm cm}$) the gas has a pressure of 
$0.1 \, {\rm dyne} \, {\rm cm}^{-2}$ and a density of $0.125 \, {\rm g} \, {\rm cm}^{-3}$. 
The results of the simulation and the analytical solution at $t=0.245\,{\rm s}$ are shown in Figure~\ref{Fig:1}.
The slight discrepancies between the analytic and numerical results are attributable to numerical diffusion associated 
with the upwind interpolation (see Stone \& Norman 1992), match the results of Stone \& Norman (1992) almost exactly,
and are entirely acceptable.

\subsection{Gravitational Collapse}

\begin{figure}
\epsscale{.35}
\plotone{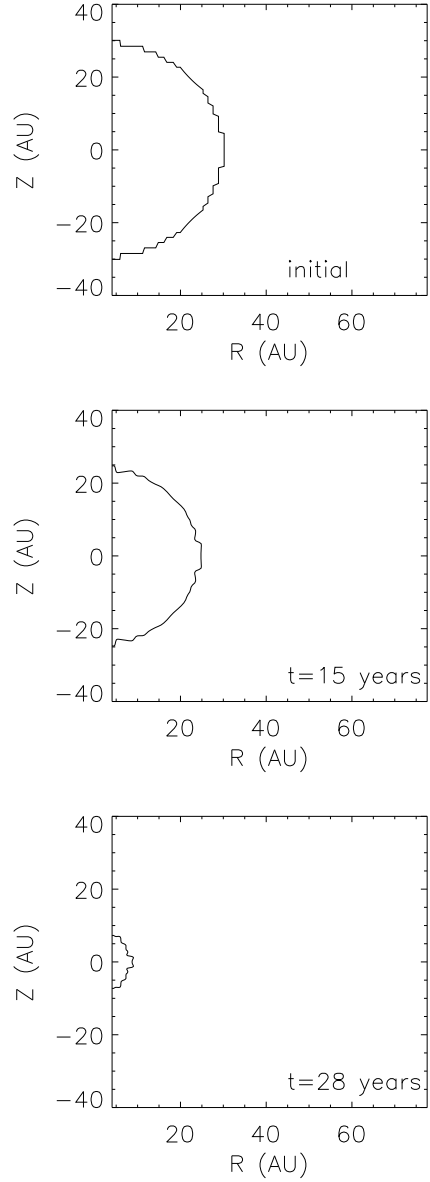}
\caption{Pressure-free collapse of a 30 AU, uniform-density clump of gas, as simulated by Perseus.
Spherical symmetry is maintained despite the cylindrical geometry and the inner boundary condition at $r = 2$ AU.}
\label{Fig:2} 
\end{figure}

As a test problem involving curvature terms, we also simulated the pressure-free gravitational collapse of a spherical 
clump of gas. 
A uniform density gas ($\rho=10^{-14}\,{\rm g\, cm^{-3}}$) was imposed everywhere within 30 AU of the star.
As stated above, the only source of gravitational acceleration in our simulations is the central protostar, 
with mass $M = 1 \, M_{\odot}$. 
The grid on which this simulation takes place has 120 zones in the $z$ direction and 80 in the $r$ direction
The free-fall timescale under the gravitational potential of a 1 $\msol$ star is 29.0 yrs.  
The results of the simulation can be seen in Figure~\ref{Fig:2}.  
After 28 years, the $30\, {\rm AU}$ clump has contracted to the edge of the computational volume.  
Spherical symmetry is maintained throughout as the gas is advected despite the presence of the inner boundary 
condition.

\subsection{Cooling}

\begin{figure}
\epsscale{0.7}
\plotone{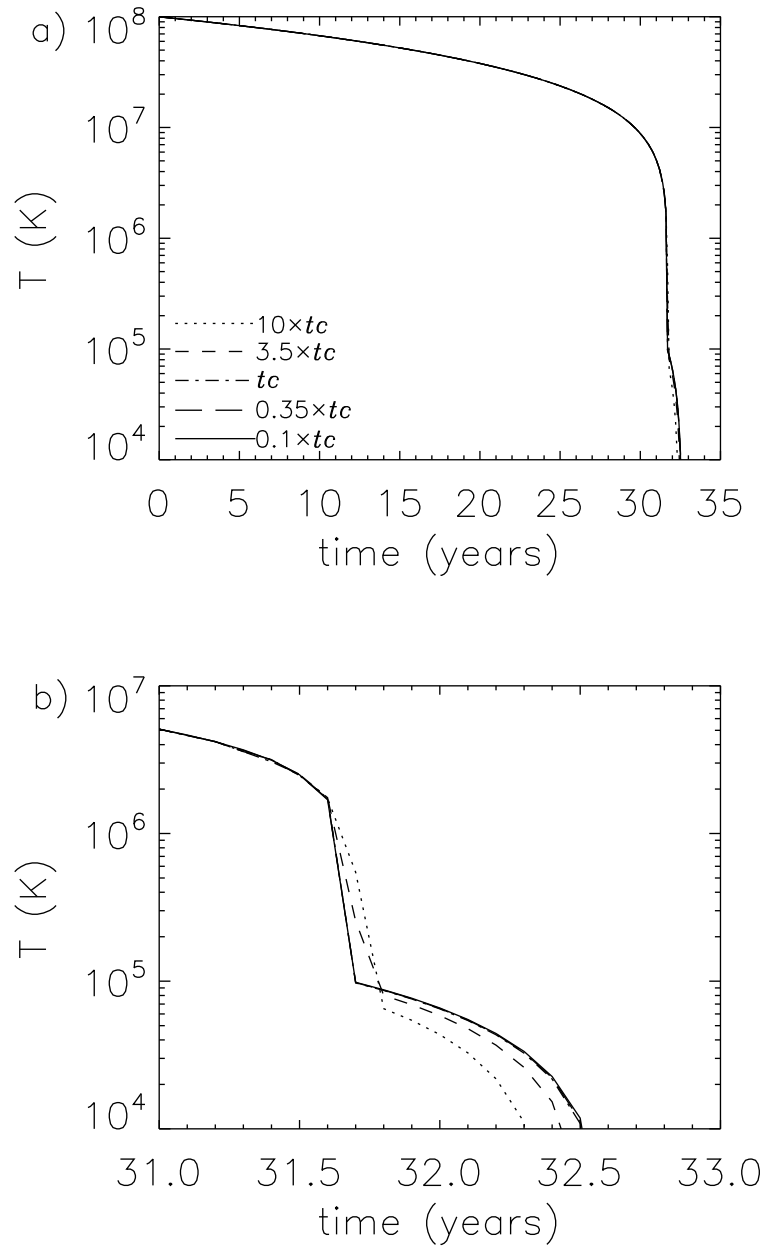}
\caption{(a) Gas temperature evolution using various cooling time steps ($t_{\rm c}$ is the time step normally used in the code) 
(b) Close-up of the time interval 31-33 years.  Convergence is achieved using $t_{\rm c}$ and higher resolution timesteps.}
\label{Fig:3} 
\end{figure}

To test the accuracy of the cooling algorithm, a simple 2D grid of 64 zones by 64 zones was set up.  
The simulation starts with gas at $T=10^{10}\, {\rm K}$.  
The temperature of the gas is followed until it reaches $T=10^{4}\, {\rm K}$.  
Simulations were run varying the cooling time step $\delta t_{\rm cool}$.   
As the cooling subroutine does not use a predictor-corrector method, decreasing the time step increases the precision.  
A range of cooling time steps, varying from 10 times what is used in the code to 0.1 times what is used in the code, were
tested.  
Since in the range of $T=10^{4}\, {\rm K}-10^{10}\, {\rm K}$, the cooling rate varies with temperature (according to 
Sutherland \& Dopita 1993), the size of the time step should affect the time evolution of the temperature.
This evolution is depicted in Figure~\ref{Fig:3}, from which one can see that $\delta t_{\rm cool}$ used in the code 
is sufficient, as using smaller time steps gives the same result.  
In addition, we can see that even the lesser precision runs give comparably good results, as the thermal time step 
of the CFL condition prevents a catastrophically rapid cooling.  
The precision of the cooling is limited by the accuracy of the cooling factors used, not the algorithm.

\subsection{``Relaxed Disk''}

Finally, we have modeled the long-term evolution of an isolated protoplanetary disk.  
To begin, a minimum-mass solar nebula disk (Hayashi et al.\ 1985) in Keplerian rotation is truncated at $30\,\rm{AU}$.  
The code then runs for 2000 years, allowing the disk to find its equilibrium configuration under gravity from the central star 
($1\,\msol$), pressure and angular momentum.  
We call this the ``relaxed disk", and use it as the initial state for the runs that follow.  
To check the long term stability of the system, we allow the relaxed disk to evolve an extra 2000 years.  
This test verifies the stability of the simulated disk against numerical effects.
In addition, using a color density, we can assess how much numerical diffusion occurs in the code.

\begin{figure}
\epsscale{.60}
\plotone{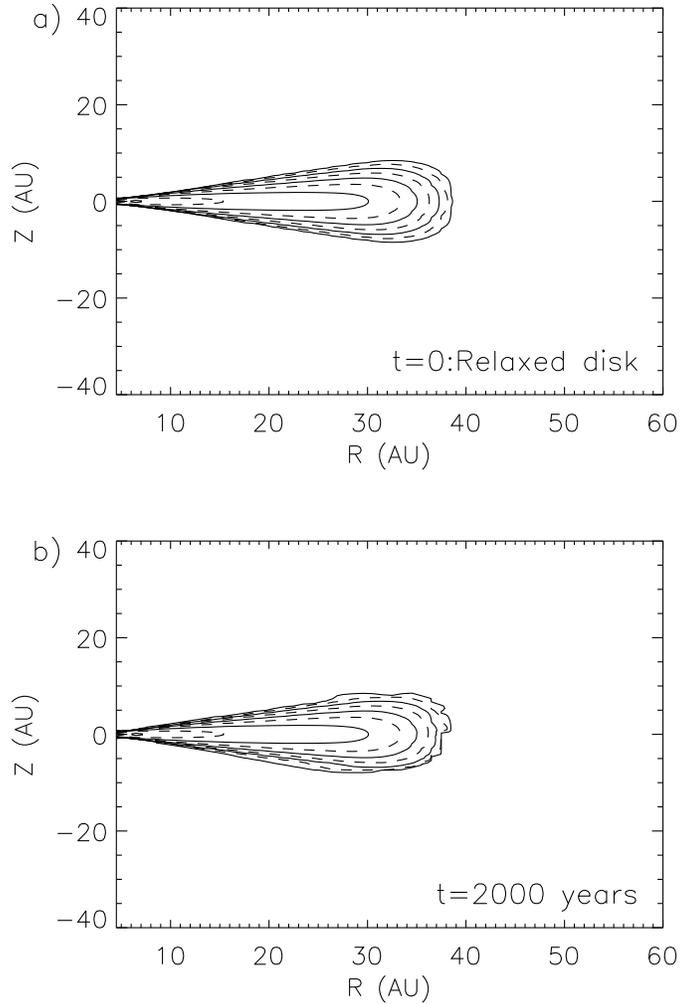}
\caption{(a) Isodensity contours of an equilibrium (``relaxed") protoplanetary disk.  Contours are spaced a factor
of 10 apart, with the outermost contour representing a density $10^{-20} \, {\rm g} \, {\rm cm}^{-3}$.   The rotating
disk has already evolved for 2000 years and is stable; this is the configuration used as the initial state for our
subsequent runs. (b) The ``relaxed" disk, after an additional 2000 years of evolution.  While some slight deformation 
of the lowest-density contours is seen, attributable to gravitational infall of surrounding gas, the disk is stable
and non-evolving over the spans of time relevant to supernova shock passage, $\approx 2000$ years.} 
\label{Fig:4}
\end{figure}

After the extra 2000 years, the disk maintains its shape, and is deformed only at its lowest isodensity contour, because 
of the gravitational infall of the surrounding gas (Figure~\ref{Fig:4}).  
Comparing this deformation to the results from the canonical run (\S 3), this is a negligible effect.
Some of the surrounding gas has accreted on the disk due to the gravitational potential of the central star.  
The color density allows us to follow the location of the accreted gas.  
After 2000 years, roughly 20\% of the accreted mass has found its way to the midplane of the disk due to the effects of 
numerical diffusion.  
Hence some numerical diffusion exists and must be considered in what follows.

\section{Canonical Case}

\begin{figure}[htbp]
\epsscale{1.0}
\begin{center}
\plotone{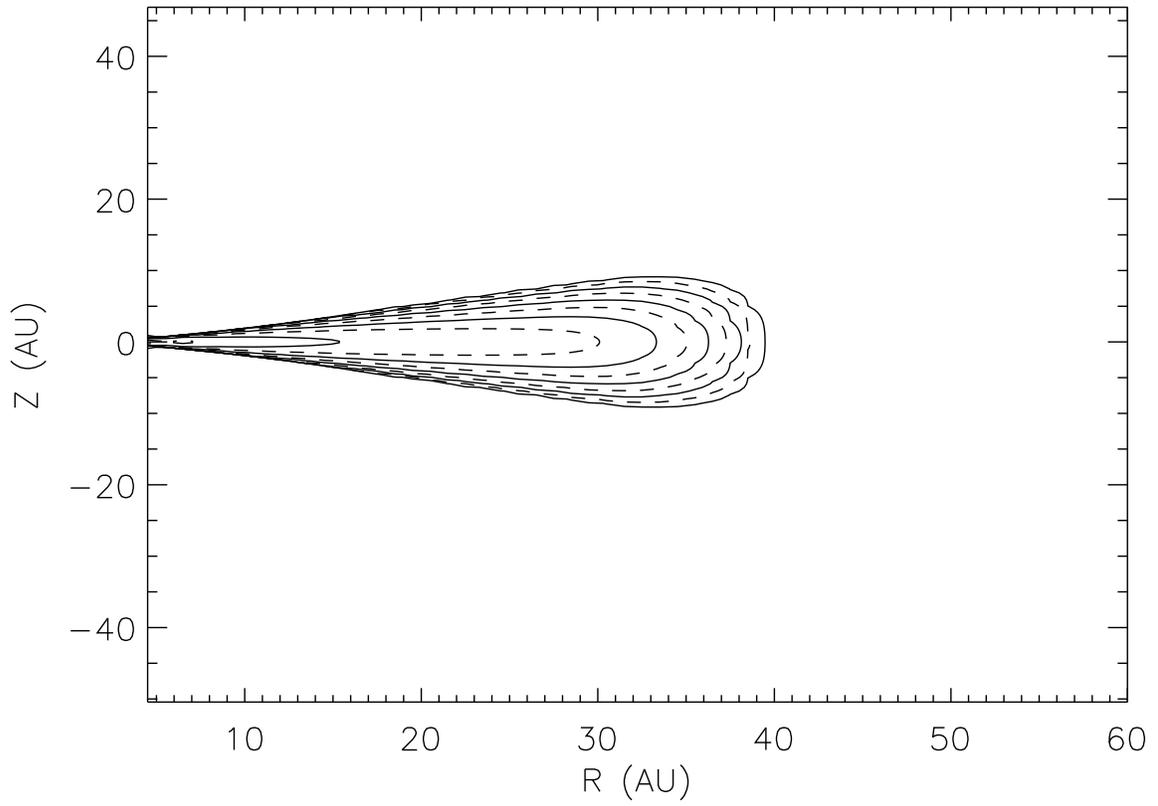}
\caption{Isodensity contours of the relaxed disk, just before impact of the supernova ejecta.  Contours are spaced 
a factor of 10 apart, with the outermost contour representing a density $10^{-21}\,{\rm g\, cm^{-3}}$.} 
\label{Fig:5}
\end{center}
\end{figure}

In this section, we adopt a particular set of parameters pertinent to the disk and the supernova, and follow the evolution 
of the disk and ejecta in some detail.  
The simulation begins with our relaxed disk (\S 2.5), seen in Figure~\ref{Fig:5}.  
Its mass is about 0.00838 $\msol$, and it extends from $4\,{\rm AU}$ to $40\,{\rm AU}$, 
the inner parts of the disk being removed to improve code performance.  
The gas density around the disk is taken to be a uniform $10\,{\rm cm^{-3}}$, which is a typical density for an $\hii$ region.
This disk has similar characteristics to those found in the Orion nebula, which have been photoevaporated down to tens
of AU by the radiation of nearby massive O stars (Johnstone, Hollenbach \&  Bally 1998). 
In setting up our disk, we have ignored the effects of the UV flash that accompanies the supernova, in which approximately 
$3 \times 10^{47} \, {\rm erg}$ of high-energy ultraviolet photons are emitted over several days (Hamuy et al.\ 1988).
The typical UV opacities of protoplanetary disk dust are $\kappa \sim 10^{2} \, {\rm cm}^{2} \, {\rm g}^{-1}$ 
(D'Allesio et al.\ 2006), so this UV energy does not penetrate below a column density 
$\sim \kappa^{-1} \sim 10^{-2} \, {\rm g} \, {\rm cm}^{-2}$. 
The gas density at the base of this layer is typically $\rho \sim 10^{-15} \, {\rm g} \, {\rm cm}^{-3}$; if the gas 
reaches temperatures $< 10^{5} \, {\rm K}$, $t_{\rm cool}$ will not exceed a few hours (\S 2.1). 
The upper layer of the disk absorbing the UV is not heated above a temperature 
$T \sim (E_{\rm UV} / 4\pi d^{2}) m_{\rm H} \kappa / k_{\rm B} \sim 10^{5} \, {\rm K}$.
Because the gas in the disk absorbs and then reradiates the energy it absorbs from the UV flash, we have ignored it.
We have also neglected low-density gas structures that are likely to have surrounded the disk, including 
photoevaporative flows and bow shocks from stellar winds, as these are beyond the scope of this paper.
It is likely that the UV flash would greatly heat this low-density gas and cause it to rapidly escape the disk
anyway.
Our ``relaxed disk" initial state is a reasonable, simplified model of the disks seen in $\hii$ 
regions before they are struck by supernova shocks.

After a stable disk is obtained, supernova ejecta are added to the system.  
The canonical simulation assumes $M_{\rm ej} = 20\, \msol$ of material was ejected isotropically by a supernova
$d = 0.3 \, {\rm pc}$ away, with an explosion kinetic energy $E_{\rm ej} = 10^{51} \, {\rm erg}$, (1 f.o.e.).
This is typical of the mass ejected by a $25 \, M_{\odot}$ progenitor star, as considered by Woosley \& Weaver (1995), 
and although more recent models show that progenitor winds are likely to reduce the ejecta mass to $< 10 \, M_{\odot}$ 
(Woosley, Heger \& Weaver 2002), we retain the larger ejecta mass as a worst-case scenario for disk survivability. 
The ejecta are assumed to explode isotropically, but with density and velocity decreasing with time.
The time dependence is taken from the scaling solutions of Matzner \& McKee (1999); in analogy to their eq.\ [1],
we define the following quantities:
\begin{eqnarray}
v_* & = & \sqrt{\frac{2 E_{\rm ej}}{M_{\rm ej}}} \nonumber \\ 
t_* & = & \frac{R_*}{v_*} \\ 
\rho_* & = & \frac{3 M_{\rm ej}}{4\pi R_*^3} \nonumber \\ 
p_* & = &\frac{3E_{\rm ej}}{4\pi R_*^3}\,, \nonumber 
\end{eqnarray}
where $R_*$ is the radius of the exploding star, taken to be $50 \, R_{\odot}$.  
The travel time from the supernova to the disk is computed as $t_{\rm trav} = d / v_*$, and is typically $\sim 100$ years.
Finally, expressions for the time dependence of velocity, density and pressure of the ejecta, are obtained for any 
given time $t$ after the shock strikes the disk: 
\begin{eqnarray}
v_{{\rm ej}}(t)    & = & v_*          \, \left( \frac{t_{\rm trav}}{t+t_{\rm trav}} \right) \nonumber \\ 
\rho_{{\rm ej}}(t) & = & \rho_* \left( \frac{t_*}{t_{\rm trav}} \right)^{3}
                               \, \left( \frac{t_{\rm trav}}{t+t_{\rm trav}} \right)^{3} \\
p_{\rm ej}(t)      & = & p_*    \left( \frac{t_*}{t_{\rm trav}} \right)^{4} 
                               \, \left( \frac{t_{\rm trav}}{t+t_{\rm trav}} \right)^{4} \nonumber 
\end{eqnarray}
We acknowledge that supernova ejecta are not distributed homogeneously within the progenitor (Matzner \& McKee 1999), 
nor are they ejected isotropically (Woosley, Heger \& Weaver 2002),  but more detailed modeling lies beyond the scope
of this paper. 
Our assumption of homologous expansion is in any case a worst-case scenario for disk survivability in that the ejecta
are front-loaded in a way that overestimates the ram pressure (C. Matzner, private communication).
As our parameter study (\S 5) shows, density and velocity variations have little influence on the results.

\begin{figure}[htbp]
\epsscale{1.0}
\begin{center}
\plotone{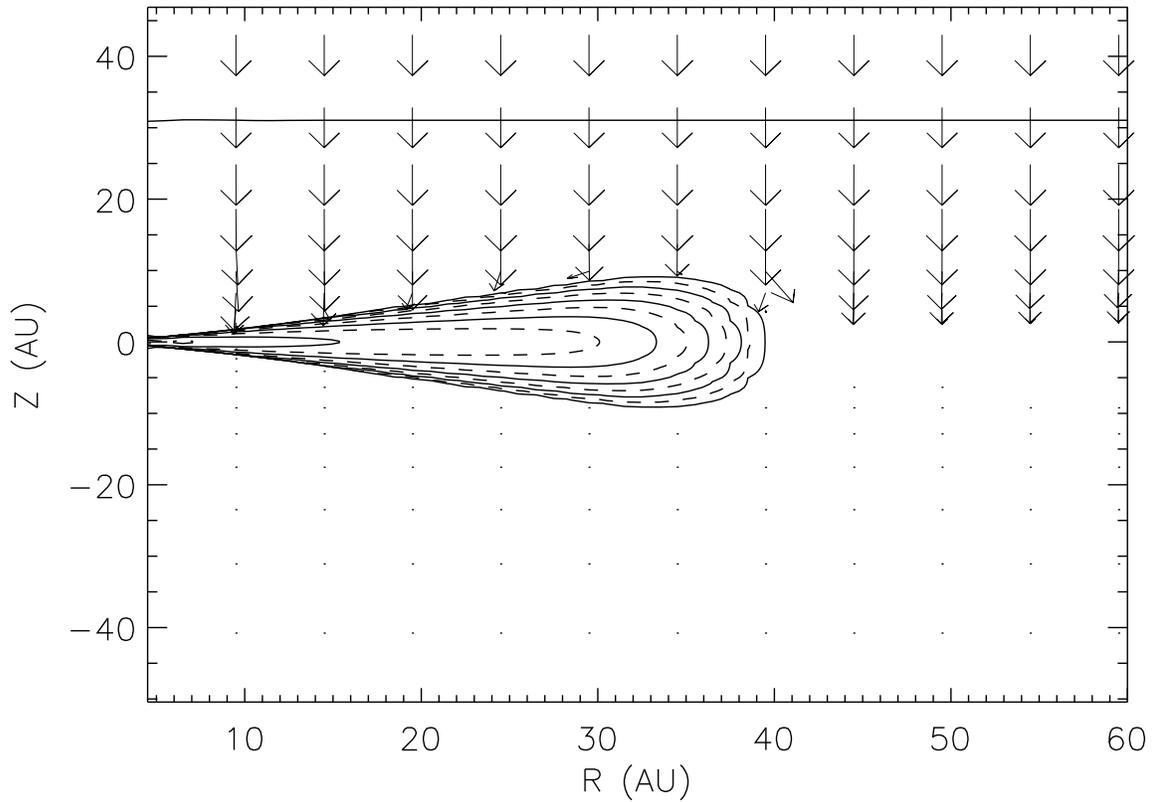}
\caption{Protoplanetary disk immediately prior to impact by the supernova shock.  Isodensity contours are as 
in Figure~\ref{Fig:4}.  Arrows represent gas velocities.  The supernova ejecta are travelling through the $\hii$
region toward the disk at about $2200 \, {\rm km} \, {\rm s}^{-1}$.} 
\label{Fig:6}
\end{center}
\end{figure}

\begin{figure}[htbp]
\epsscale{1.0}
\begin{center}
\plotone{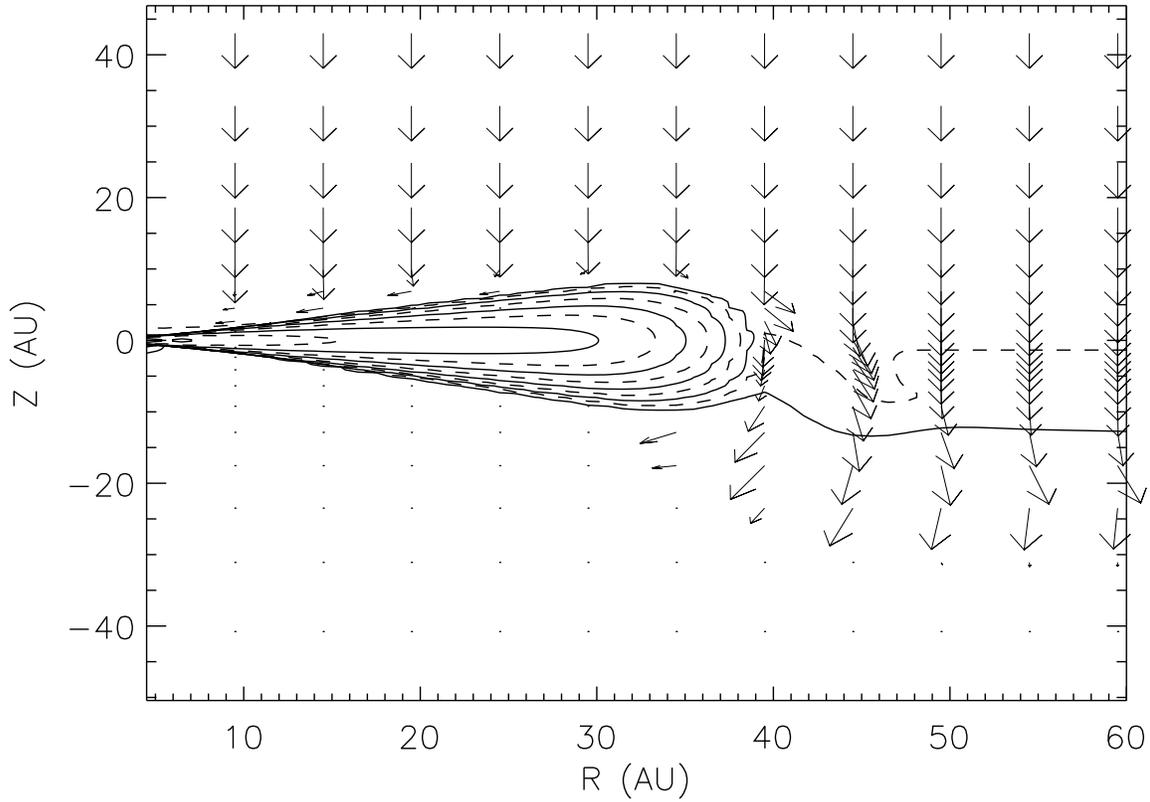}
\caption{Protoplanetary disk 0.05 years after being first hit by supernova ejecta.  As the supernova shock sweeps 
around the disk edge, it snowplows the low-density disk material with it, but the shock stalls in the high-density gas 
in the disk proper.  Isodensity contours and velocity vectors as in Figure~\ref{Fig:5}.}
\label{Fig:7}
\end{center}
\end{figure}

\begin{figure}[htbp]
\epsscale{1.0}
\begin{center}
\plotone{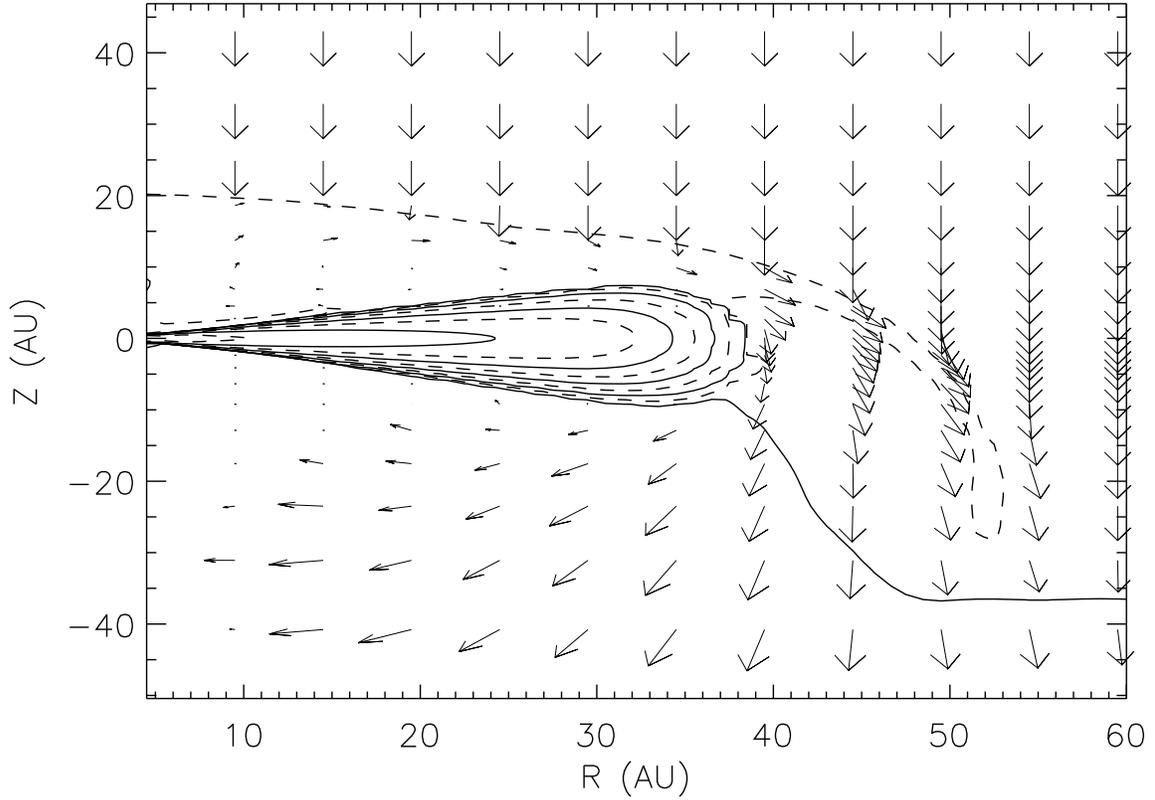}
\caption{Protoplanetary disk 0.1 years after first being hit by supernova ejecta.  As the pressure increases on the 
side of the disk facing the ejecta, a reverse shock forms, visible as the outermost (dashed) isodensity contour.
Isodensity contours and velocity vectors as in Figure~\ref{Fig:5}.} 
\label{Fig:8}
\end{center}
\end{figure}

\begin{figure}[htbp]
\epsscale{1.0}
\begin{center}
\plotone{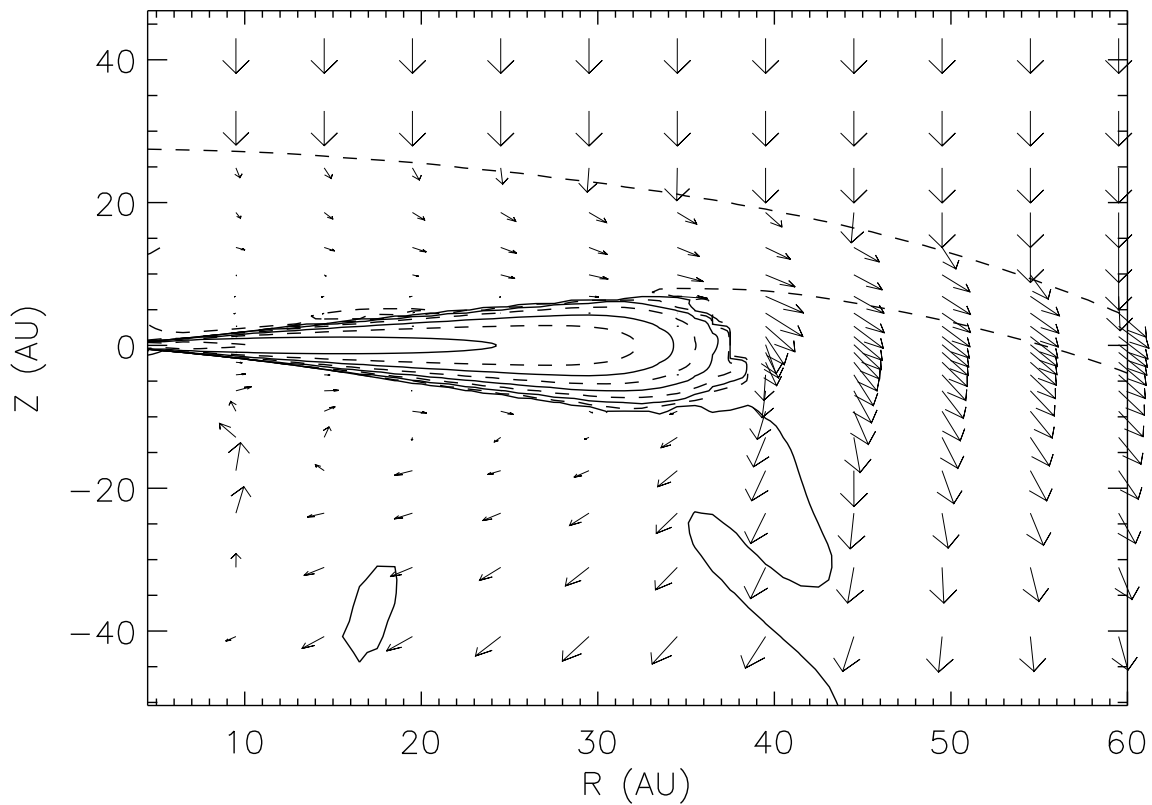}
\caption{Protoplanetary disk 0.3 years after first being hit by supernova ejecta.  The reverse shock, visible as the 
outermost (dashed) contour, has stalled and formed a bow shock.  The bow shock deflects incoming gas around the disk,
which is effectively protected in a high-pressure ``bubble" of gas.  Isodensity contours and velocity vectors as in 
Figure~\ref{Fig:5}.} 
\label{Fig:9}
\end{center}
\end{figure}

The incoming ejecta and the shock they create while propagating through the low-density gas of the $\hii$ region can 
be seen in Figure~\ref{Fig:6}.
When the shock reaches the disk, the lower-density outer edges are swept away, as the ram pressure of the ejecta is much 
higher than the gas pressure in those areas. 
However, the shock stalls at the higher density areas of the disk, as the gas pressure is higher there.   
A snapshot of the stalling shock can be seen in Figure~\ref{Fig:7}.
As the ejecta hit the disk, they shock and thermalize, heating the gas on the upper layers of the disk.  
This increases the pressure in that area, causing a reverse shock to propagate into the incoming ejecta.  
The reverse shock will eventually stall, forming a bow shock around the disk (Figures~\ref{Fig:8} and~\ref{Fig:9}).  
Roughly 4 months have passed between the initial contact and the formation of the bow shock.

\begin{figure}[htbp]
\epsscale{1.0}
\begin{center}
\plotone{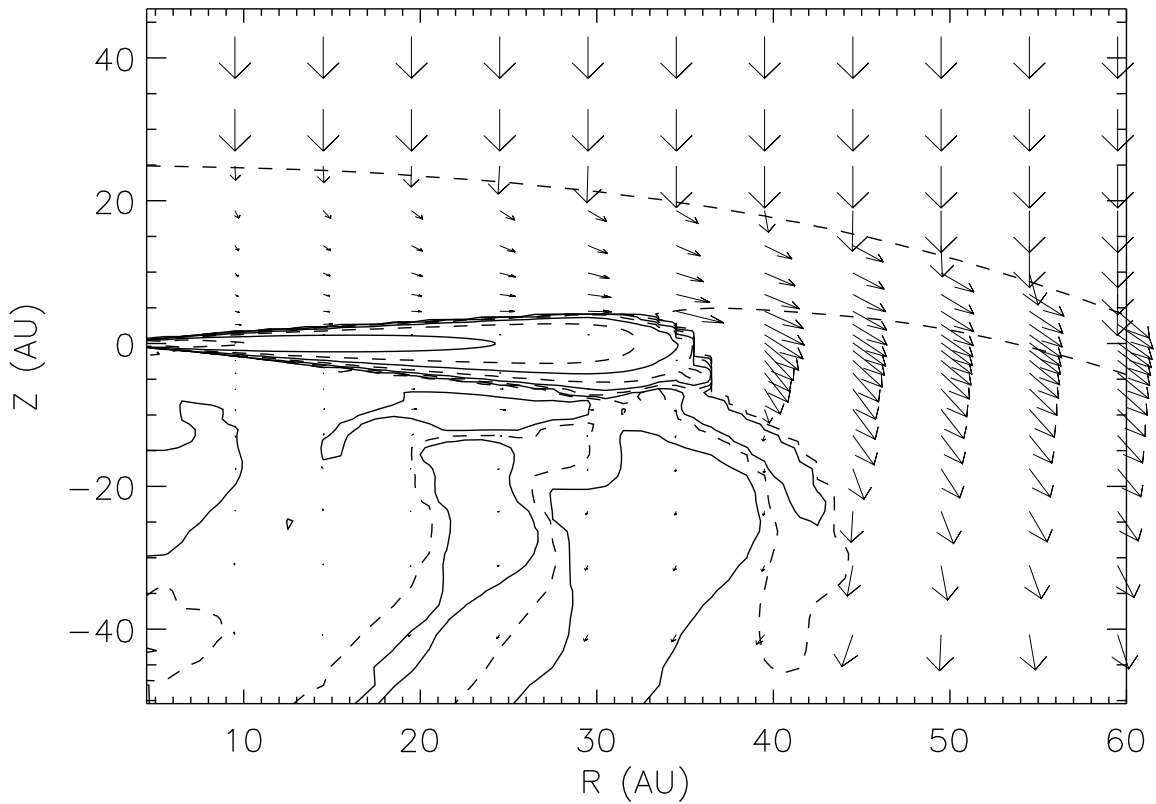}
\caption{Protoplanetary disk 4 years after first being hit by supernova ejecta.  Gas is being stripped from the disk 
(e.g., the clump between $R = 35$ and 45 AU).  Gas stripped from the top of the disk either is entrained in the flow of
ejecta and escapes the simulation domain, or flows under the disk and falls back onto it. 
Isodensity contours and velocity vectors as in Figure~\ref{Fig:5}.} 
\label{Fig:10}
\end{center}
\end{figure}

Some stripping of the low density gas at the disk's edge ($> 30$ AU) may occur as the supernova ejecta is deflected 
around it, due primarily to the ram pressure of the ejecta.
As the stripped gas is removed from the top and the sides of the disk, it either is snowplowed away from the disk 
if enough momentum has been imparted to it, or it is pushed behind the disk, where it can fall back onto it 
(Figure~\ref{Fig:10}).
In addition to stripping the outer layers of the disk, the pressure of the thermalized shocked gas will compress the disk to a 
smaller size; although they do not destroy the disk, the ejecta do temporarily deform the disk considerably.
Figure~\ref{Fig:11} shows the effect of the pressure on the disk, which has been reduced in thickness and has shrunk to a 
radius of 30 AU.  
The extra external pressure effectively aids gravity and allows the gas to orbit at a smaller radius with the same angular 
momentum.
As the ejecta is deflected across the top edge of the disk, some mixing between the disk gas and the ejecta may occur 
through Kelvin-Helmholtz instabilities.  
Figure~\ref{Fig:12} shows a close up of the disk where a Kelvin-Helmholtz roll is occurring at the boundary between the 
disk and the flowing ejecta. 
In addition, some ejecta mixed in with the stripped material under the disk might also accrete onto the disk.
As time goes by and slower ejecta hit the disk, the ram pressure affecting the disk diminishes, and the disk slowly 
returns to its original state, recovering almost completely after 2000 years (Figure~\ref{Fig:13}).

The exchange of material between the disk and the ejecta is mediated through the ejecta-disk interface, which
in our simulations is only moderately well resolved.  
As discussed in \S 4, the numerical resolution will affect how well we quantify both the destruction of the disk and 
the mixing of ejecta into the disk.
In the canonical run, at least, disk destruction and gas mixing are minimal. 
Although some stripping has occurred while the disk was being hit by the ejecta, it has lost less than 0.1\% of its mass.  
The final disk mass, computed from the zones where the density is greater than $100\,{\rm cm^{-3}}$, remains roughly at 
0.00838 $\msol$.
Some of the ejecta have also been mixed into the disk, but only with very low efficiency.
A $30\,\rm{AU}$ disk sitting $0.3\,{\rm pc}$ from the supernova intercepts roughly one part in $1.7 \times 10^7$ of the total 
ejecta from the supernova, assuming isotropic ejecta distribution.  
For 20 $\msol$ of ejecta, this corresponds to roughly $1.18 \times 10^{-6}\, \msol$ intercepted.  
At the end of the simulation, we find only $1.48 \times 10^{-8}\, \msol$ of supernova ejecta was injected in the disk, for 
an injection efficiency of about 1.3\%.  
Some of the injected material could be attributed to numerical diffusion between the outer parts of the disk and the inner 
layers: as seen in \S 2.5, Perseus is diffusive over long periods of time.  
However, the distribution of the colored mass is qualitatively different from that obtained from a simple numerical diffusion 
process.  
Figure~\ref{Fig:14} compares the percentage of colored mass within a given isodensity contour for the canonical case and the 
relaxed disk simulation of \S 2.5, at a time 500 years after the beginning of each of these simulations.  
From this graph, it  is clear that the process that injects the supernova ejecta is not simply numerical diffusion, as it is much 
more efficient at injecting material deep within the disk.  
The post-shock pressure of the ejecta gas, 100 years after initial contact, when its forward progession in the disk has 
stalled is $\sim 2 \rho_{\rm ej} v_{\rm ej}^2 / (\gamma +1) = 2.8 \times 10^{-5}\, {\rm dyne} \, {\rm cm}^{-2}$.
(After 100 years, $\rho_{\rm ej} = 2.2 \times 10^{-21}\, {\rm g} \, {\rm cm}^{-3}$ and 
$v_{\rm ej} = 1300 \, {\rm km} \, {\rm s}^{-1}$.)
The shock stalls where the post-shock pressure is comparable to the disk pressure $\sim \rho k_B T/\bar{m}$.  
Hence at $20\,{\rm AU}$, where the temperature of the disk is $T \approx 30\,{\rm K}$, the shock stalls at the isodensity
contour $\sim 1.5 \times 10^{-14}\, {\rm g\, cm^{-3}}$.  
As about half of the color mass is mixed to just this depth, this is further evidence that the color field in the disk 
represents a real physical mixing.  

\section{Numerical Resolution}

The results of canonical run show many similarities to related problems that have been studied extensively
in the literature.  
The interaction of a supernova shock with a protoplanetary disk resembles the interaction of a shock with 
a molecular cloud, as modeled by Nittmann et al.\ (1982), Bedogni \& Woodward (1990), 
Klein, McKee \& Colella (1994; hereafter KMC), Mac Low et al.\ (1994), Xu \& Stone (1995), Orlando et al.\ (2005)
and Nakamura et al.\ (2006). 
Especially in Nakamura et al.\ (2006), the numerical resolutions achieved in these simulations are state-of-the-art, 
reaching several $\times 10^{3}$ zones per axis. 
In those simulations, as in our canonical run, the evolution is dominated by two physical effects:
the transfer of momentum to the cloud or disk; and the onset of Kelvin-Helmholtz (KH) instabilities
that fragment and strip gas from the cloud or disk. 
KH instabilities are the most difficult aspect of either simulation to model, because there is no practical
lower limit to the lengthscales on which KH instabilities operate (they are only suppressed at scales smaller 
than the sheared surface). 
Increasing the numerical resolution generally reveals increasingly small-scale structure at the interface 
between the shock and the cloud or disk (see Figure 1 of Mac Low et al.\ 1994). 
The numerical resolution in our canonical run is about 100 zones per axis; more specifically, there are about 
26 zones in one disk radius (of 30 AU), and about 20 zones across two scale heights of the disk (one scale-height 
being about 2 AU at 20 AU).
Our highest-resolution run used about 50 zones along the radius of the disk, and placed about 30 zones across
the disk vertically. 
In the notation of KMC, then, our simulations employ about 20-30 zones per cloud radius, a factor of 3 lower than
the resolutions of 100 zones per cloud radius argued by Nakamura et al.\ (2006) to be necessary to resolve 
the hydrodynamics of a shock hitting a molecular cloud. 

Higher numerical resolutions are difficult to achieve; unlike the case of a supernova shock with speed 
$\sim 2000 \, {\rm km} \, {\rm s}^{-1}$ striking a molecular cloud with radius of 1 pc, our simulations deal with 
a shock with the same speed striking an object whose intrinsic lengthscale is $\sim 0.1 \, {\rm AU}$. 
Satisfying our CFL condition requires us to use timesteps that are only $\sim 10^{3} \, {\rm s}$, four
orders of magnitude smaller than the timesteps needed for the case of a molecular cloud.  
This and other factors conspire to make simulations of a shock striking a protoplanetary disk about 100 times
more computationally intensive than the case of a shock striking a molecular cloud.
Due to the numerous lengthscales in the problem imposed by the star's gravity and the rotation of the disk, 
it is not possible to run the simulations at low Mach numbers and then scale the results to higher Mach numbers. 
We intend to create a parallelized version of Perseus to run on a computer cluster in the near future, but until
then, our numerical resolution cannot match that of simulations of shocks interacting with molecular clouds.
This begs the question, if our resolution is not as good as has been achieved by others, is it good enough? 

To quantify what numerical resolutions are sufficient, we examine the physics of a shock interacting with
a molecular cloud, and review the convergence studies of the same undertaken by previous authors. 
In the most well-known simulations (Nittmann et al. 1982; KMC; Mac Low et al.\ 1994; Nakamura et al.\ 2006), 
it is assumed that a low-density molecular cloud with no gravity or magnetic fields is exposed to a steady shock.
The shock collides with the cloud, producing a reverse shock that develops into a bow shock; a shock propagates
through the cloud, passing through it in a ``cloud-crushing" time $t_{\rm cc}$. 
The cloud is accelerated, but as long as a velocity difference between the high-velocity gas and the cloud exists, 
KH instabilities grow that create fragments with significant velocity dispersions, $\sim 10\%$ of the 
shock speed (Nakamura et al.\ 2006). 
Cloud destruction takes place before the cloud is fully accelerated, and the cloud is 
effectively fragmented in a few $\times \, t_{\rm cc}$ before the velocity difference diminishes.
These fragments are not gravitationally bound to the cloud and easily escape.
As long as the shock remains steady for a few $\times \, t_{\rm cc}$, it is inevitable that the cloud is destroyed. 

As KH instabilities are what fragment the cloud and accelerate the fragments, it is important to model them
carefully, with numerical resolution as high as can be achieved. 
KMC stated in their abstract and throughout their paper that 100 zones per cloud radius were required for 
``accurate results"; however, all definitions of what was meant by ``accurate", or what were the physically relevant
``results" were deferred to a future ``Paper II". 
A companion paper by Mac Low et al.\ (1994) referred to the same Paper II and repeated the claim that 100 zones
per axis were required. 
Nakamura et al.\ (2006), published this year, appears to be the Paper II that reports the relevant convergence study
and quantifies what is meant by accurate results. 
Global quantities, including the morphology of the cloud, its forward mass and momentum, and the velocity dispersions
of cloud fragments, were defined and calculated at various levels of numerical resolution.  
These were then compared to the same quantities calculated using the highest achievable resolutions, about 500
zones per cloud radius (over 1000 zones per axis).
The quantities slowest to converge with higher numerical resolution were the velocity dispersions, probably, they
claim, because these quantities are so sensitive to the hydrodynamics at shocks and contact discontinuities where 
the code becomes first-order accurate only. 
The velocity dispersions converged to within 10\% of the highest-resolution values only when at least 100 zones per 
cloud radius were used. 
For this single arbitrary reason, Nakamura et al.\ (2006) claimed numerical resolutions of 100 zones per cloud radius 
were necessary. 
We note, however, that the other quantities to do with cloud morphology and momentum were found to converge much 
more readily; according to Figure 1 of Nakamura et al.\ (2006), numerical resolutions of only 30 zones per cloud 
radius are sufficient to yield values within 10\% of the values found in the highest-resolution simulations. 
And although the velocity dispersions are not so well converged at 30 zones per cloud radius, even then the errors 
do not exceed a factor of 2. 
Assuming that the problem we have investigated is similar enough to that investigated by Nakamura et al.\ (2006) so
that their convergence study could be applied to our problem, we would conclude that even our canonical run is 
sufficiently resolving relevant physical quantities, the one possible exception being the velocities of fragments 
generated by KH instabilities, where the errors could be a factor of 2. 

Of course, the problem we have investigated, a supernova shock striking a protoplanetary disk, is different in 
four very important ways from the cases considered by KMC, Mac Low et al.\ (1994) and Nakamura et al.\ (2006).
The most important fundamental difference is that the disk is gravitationally bound to the central protostar. 
Thus, even if gas is accelerated to supersonic speeds $\sim 10 \, {\rm km} \, {\rm s}^{-1}$, it is not guaranteed to
escape the star.
Second, the densities of gas in the disk, $\rho_{\rm disk}$, are significantly higher than the density in the gas 
colliding with the disk, $\rho_{\rm ej}$. 
In the notation of KMC, $\chi = \rho_{\rm disk} / \rho_{\rm ej}$. 
Because the disk density is not uniform, no single value of $\chi$ applies, but if $\chi$ is understood to 
refer to different parcels of disk gas, $\chi$ would vary from $10^{4}$ to over $10^{8}$. 
This affects the magnitudes of certain variables (see, e.g., Figure 17 of KMC regarding mix fractions), but also 
{\it qualitatively} alters the problem: the densities and pressures in the disk are so high that the supernova shock 
cannot cross through the disk, instead stalling at several scale heights above the disk.
Unlike the case of a shock shredding a molecular cloud, the cloud-crushing timescale $t_{\rm cc}$ is not even
a relevant quantity for our calculations. 
The third difference is that shocks cannot remain non-radiative when gas is as dense as it is near the disk.
Using $\rho = 10^{-14} \, {\rm g} \, {\rm cm}^{-3}$ and $\Lambda = 10^{-24} \, {\rm erg} \, {\rm cm}^{3} \, {\rm s}^{-1}$,
$t_{\rm cool}$ is only a few hours, and shocks in the disk are effectively isothermal. 
Shocks propagating into the disk therefore stall at somewhat higher locations above the disk than they would have 
if they were adiabatic. 
Finally, the fourth fundamental difference between our simulations and those investigated in KMC, Mac Low et al.\ (1994)
and Nakamura et al.\ (2006) is that we do not assume steady shocks.  
For supernova shocks striking protoplanetary disks about 0.3 pc away, the most intense effects are felt only for
a time $\sim 10^{2}$ years, and after only 2000 years the shock has for all purposes passed. 
There are limits, therefore, to the energy and momentum that can be delivered to the disk. 
Very much unlike the case of a steady, non-radiative shock striking a low-density, gravitationally unbound molecular
cloud, where ultimately destruction of the cloud is inevitable, many factors contribute to the survivability of 
protoplanetary disks struck by supernova shocks. 

This conclusion is borne out by a resolution study we have conducted that shows that the vertical momentum delivered to 
the disk is certainly too small to destroy it, and that we are not significantly underresolving the KH instabilities at
the top of the disk. 
Using the parameters of our canonical case, we have conducted 6 simulations with different numerical resolutions.
The resolutions range from truly awful, with only 8 zones in the radial direction ($\Delta r = 10 \, {\rm AU}$) and 
18 zones in the vertical direction (with $\Delta z = 1 \, {\rm AU}$ at the midplane, barely sufficient to resolve a 
scale height), to our canonical run (76 x 120), to one high-resolution run with 152 radial zones
($\Delta r = 0.5 \, {\rm AU}$) and 240 vertical zones ($\Delta z = 0.13 \, {\rm AU}$ at the midplane).
On an Apple G5 desktop with two 2.0-GHz processors, these simulations took from less than a day to 80 days to run. 
To test for convergence, we calculated several global quanities $Q$, including: the density-weighted cloud radius, $a$;
the density-weighted cloud thickness, $c$; the density-weighted vertical velocity, $\vz$; the density-weighted 
velocity dispersion in $r$, $\delta v_{r}$; the density-weighted velocity dispersion in $z$, $\vz$; as well 
as the mass of ejecta injected into the disk, $M_{\rm inj}$.
Except for the last quantity, these are defined exactly as in Nakamura et al.\ (2006), but using a density threshold 
corresponding to $100 \, {\rm cm}^{-3}$.
Each global quantity was measured at a time 500 years into each simulation. 
We define each global quantity $Q$ as a function of numerical resolution $n$, where $n$ is the geometric mean of the
number of zones along each axis, which ranges from 12 to 191. 
To compare to the resolutions of KMC, one must divide this number by about 3 to get the number of zones per 
``cloud radius" (two scale heights at 20 AU) in the vertical direction, and divide by about 2 to get the number of 
zones per cloud radius in the radial direction. 
The convergence is measured by computing $\left| Q(n) - Q(n_{\rm max}) \right| / Q(n_{\rm max})$, where $n_{\rm max} = 191$ 
corresponds to our highest resolution case.
In Figure~\ref{Fig:15} we plot each quantity $Q(n)$ as a function of resolution $n$ (except $\vz$).
All of the quantities have converged to within 10\%, the criterion imposed by Nakamura et al.\ (2006) as signifying 
adequate convergence.  
It is significant that $\delta v_{r}$ has converged to within 10\%, because this is the quantity relevant to 
disk destruction by KH instabilities. 
Material is stripped from the disk only if supersonic gas streaming radially above the top of the disk can 
generate KH instabilities and fragments of gas that can then be accelerated radially to escape velocities.
If we were underresolving this layer significantly, one would expect large differences in $\delta v_{r}$ as
the resolution was increased, but instead this quantity has converged.  
Higher-resolution simulations are likely to reveal smaller-scale KH instabilities and perhaps more stripping
of the top of the disk, but not an order of mangitude more. 

The convergence of $\vz$ with resolution is handled differently because unlike the other quantities, $\vz$
can vanish at certain times. 
The disk absorbs the momentum of the ejecta and is pushed downward, but unlike the case of an isolated molecular
cloud, the disk feels a restoring force from the gravity of the central star.  
The disk then undergoes damped vertical oscillations about the origin as it collides with incoming ejecta at
lower and lower speeds. 
This behavior is illustrated by the time-dependence of $\vz$, shown in Figure~\ref{Fig:16} for two numerical resolutions, 
our canonical run ($n = 95$) and our highest-resolution run ($n = 191$).
Figure~\ref{Fig:16} shows that the vertical velocity of the disk oscillates about zero, but with an amplitude 
$\sim 0.1 \, {\rm km} \, {\rm s}^{-1}$.
The time-average of this amplitude can be quantified by $\left( \overline{\vz}^{2} - \overline{<v_{z}>^{2}} \right)^{1/2}$,
where the bar represents an average over time; the result is $825 \, {\rm cm} \, {\rm s}^{-1}$ for the highest-resolution
run and is only 2\% smaller for the canonical resolution. 
The difference between the two runs is generally much smaller than this; except for a few times around $t = 150$ yr, and 
$t = 300$ yr, when the discrepancies approach 30\%, the agreement between the two resolutions is within 10\%.
The time-averaged dispersion of the amplitude of the difference (defined as above for $\vz$ itself) is only
$12.0 \, {\rm cm} \, {\rm s}^{-1}$, which is only 1.5\% of the value for $\vz$ itself. 
Taking a time average of $\left| \vz_{95} - \vz_{191} \right| / \left| \vz_{191} \right|$ yields 8.7\%.
We therefore claim convergence at about the 10\% level for $\vz$ as well. 

Using these velocities, we also note here that the neglect of the star's motion is entirely justified. 
The amplitude of $\vz$ is entirely understandable as reflecting the momentum delivered to the disk by the supernova 
ejecta, which is $\sim 20 \, M_{\odot} \, (\pi R_{\rm disk}^{2} / 4\pi d^{2}) \, V_{\rm ej} \sim 
10^{-3} \, M_{\odot} \, {\rm km} \, {\rm s}^{-1}$, and which should yield a disk velocity $\sim 0.1 \, {\rm km} \, {\rm s}^{-1}$.
The period of oscillation is about 150 years, which is consistent with most of this momentum being delivered to the 
outer reaches of the disk from 25 to 30 AU where the orbital periods are 125 to 165 years.
These velocities are completely unaffected by the neglected velocity of the central star, whose
mass is 120 times greater than the disk's mass.  
If the central star, with mass $\sim 1 \, M_{\odot}$, had been allowed to absorb the ejecta's momentum, it would only 
move at $\sim 100 \, {\rm cm} \, {\rm s}^{-1}$ and be displaced at most 0.4 AU after 2000 years. 
This neglected velocity, is much smaller than all other relevant velocities in the problem, including
$\left| \vz \right| \sim 800 \, {\rm cm} \, {\rm s}^{-1}$, as well as the escape velocities 
($\sim 10 \, {\rm km} \, {\rm s}^{-1}$), the velocities of gas flowing over the disk ($\sim 10^{2} \, {\rm km} \, {\rm s}^{-1}$),
and of course the shock speeds ($\sim 10^{3} \, {\rm km} \, {\rm s}^{-1}$). 

Our analysis shows that we have reached adequate convergence with our canonical numerical resolution ($n = 95$).
We observe KH instabilities in all of our simulations (except $n = 12$), and we see the role they play in stripping
the disk and mixing ejecta gas into it. 
We are therefore confident that we are adequately resolving these hydrodynamic features; nevertheless, we now consider 
a worst-case scenario in which we KH instabilities can strip the disk with 100\% efficiency where they act, and ask
how much mass the disk could possibly lose under such conditions. 

Supernova ejecta that has passed through the bow shock and strikes the disk necessarily stalls where the gas pressure
in the disk exceeds the ram pressure of the ejecta.  
Below this level, the momentum of the ejecta is transferred not as a shock but as a pressure (sound) wave.  
Gas motions below this level are subsonic. 
Note that this is drastically different from the case of an isolated molecular cloud as studied by KMC and others;
the high pressure in the disk is maintained only because of the gravitational pull of the central star. 

The location where the incoming ejecta stall is easily found.
Assuming the vertical isothermal minimum-mass solar nebula disk of Hayashi et al.\ (1985), 
the gas density varies as 
$\rho(r,z) = 1.4 \times 10^{-9} \, (r / 1 \, {\rm AU})^{-21/8} \, \exp( -z^{2}/2H^{2}) \, {\rm g} \, {\rm cm}^{-3}$,
where $H = c_{\rm s} / \Omega$, $c_{\rm s}$ is the sound speed and $\Omega$ is the Keplerian orbital frequency. 
Using the maximum density and velocity of the incoming ejecta ($\rho_{\rm ej} = 1.2 \times 10^{-20} \, {\rm g} \, {\rm cm}^{-3}$
and $V_{\rm ej} = 2200 \, {\rm km} \, {\rm s}^{-1}$),  the ram pressure of the shock striking the disk
does not exceed $p_{\rm ram} = \rho_{\rm ej} V_{\rm ej}^{2} / 4 = 1.5 \times 10^{-4} \, {\rm dyne} \, {\rm cm}^{-2}$ 
(the factor of 1/4 arises because the gas must pass through the bow shock before it strikes the disk).
At 10 AU the pressure in the disk, $\rho c_{\rm s}^{2}$, exceeds the ram pressure at $z = 2.7 H$, and at 20 AU
the ejecta stall at $z = 1.7 H$; the gas densities at these locations are $\approx 10^{-13} \, {\rm g} \, {\rm cm}^{-3}$.
At later times, the ejecta stall even higher above the disk, because $p_{\rm ram} \propto t^{-5}$ (cf. eq.\ [11]).
For example, at $t = 100 \, {\rm yr}$, the ram pressure drops below $1 \times 10^{-5} \, {\rm dyne} \, {\rm cm}^{-2}$,
and the ejecta stall above  $z = 3.6 H$ (10 AU) and $z = 2.9 H$ (20 AU).

The column density above a height $z$ in a vertically isothermal disk is easily found to be 
$\Sigma(>z) \approx \rho(z) H^{2} / z = p(z) / (\Omega^{2} z)$.
Integrating over radius, the total amount of disk gas that ever comes into contact with ejecta is (approximating
$z = 2 H$):
\begin{equation}
M_{\rm ss} = \int_{0}^{R_{\rm d}} 2\pi r \frac{ p_{\rm ram} r^{3} }{ G M_{\odot} z } \, {\rm d}r \approx 
\frac{\pi p_{\rm ram} R_{\rm d}^{4}}{4 G M_{\odot}}.
\end{equation} 
Using a disk radius $R_{\rm d} = 30 \, {\rm AU}$, the maximum amount of disk gas that is actually exposed to 
a shock at any time is only $1.5 \times 10^{-5} \, M_{\odot}$, or 0.2\% of the disk mass.
This fraction decreases with time as $p_{\rm ram} \propto t^{-5}$ (eq.\ [11]); the integral over time of $p_{\rm ram}$ 
is $p_{\rm ram}(t=0) \times t_{\rm trav} / 4$. 
The ram pressure drops so quickly, that effectively ejecta interact with this uppermost 0.2\% of the disk mass only
for about 30 years.
This is equivalent to one orbital timescale at 10 AU, so the amount of disk gas that is able to mix or otherwise 
interact with the ejecta hitting the upper layers of the disk is very small, probably a few percent at most. 
As for KH instabilities, they are initiated when the Richardson number drops below a critical value, when
\begin{equation}
{\rm Ri} = \left( \frac{1}{\rho} \frac{\partial \rho}{\partial z} \right) \,
            \frac{ g }{ (\partial U / \partial z)^{2} } < \frac{1}{4},
\end{equation} 
where $g = -\Omega^{2} z$ is the vertical gravitational acceleration, $\Omega$ is the Keplerian orbital frequency,
and $(\partial U / \partial z)$ is the velocity gradient at the top of the disk. 
Below the stall point, all gas motions are subsonic and the velocity gradient would have to be execptionally steep,
with an unreasonably thin shear layer thickness, $\ltsimeq H / 10$, to initiate KH instabilities.
Mixing of ejecta into the disk is quite effective above where the shock stalls, as illustrated by Figure~\ref{Fig:14};
it is in these same layers (experiencing supersonic velocities) that we expect that KH instabilities to occur, but
again $\ltsimeq 1\%$ of the disk mass can be expected to interact with these layers. 

To summarize, our numerical simulations are run at a lower simulation (by a factor of about 3) than has been claimed 
necessary to study the interaction of steady shocks with gravitationally unbound molecular clouds, but the drastically
different physics of the problem studied here as allowed us to achieve numerical convergence and allowed us to 
reach meaningful conclusions.
Our global quantities have converged to within 10\%, the same criterion used by Nakamura et al.\ (2006) to claim
convergence. 
The problem is so different because the disk is tightly gravitationally bound to the star and the supernova shock
is of finite duration.  
The high pressure in the disk makes the concept of a cloud-crushing time meaningless, because the ejecta stall 
before they drive through even 1\% of the disk gas.
Rather than a sharp interface between the ejecta and the disk, the two interact via sound waves within the disk,
which entails smoother gradients.
While we do resolve KH instabilities in this interface, we allow that we may be underresolving this layer; but even
if we are, this will not affect our conclusions regarding the disk survival or the amount of gas mixed into the disk.
This is because we already find that mass is stripped from the disk and ejecta are mixed into the disk very
effectively (see Figure~\ref{Fig:14}) above the layer where the ejecta stall, and below this layer mixing 
is much less efficient and all the gas is subsonic and bound to the star.
It is inevitable that mass loss and mixing of ejecta should be only at the $\sim 1\%$ level.
Similar studies using higher numerical resolutions are likely to reveal more detailed structures at the disk-ejecta
interface, but it is doubtful that more than a few percent of the disk mass can be mixed-in ejecta, and it is even
more doubtful that even 1\% of the disk mass can be lost. 
We therefore have sufficient confidence in our canonical resolution to use it to test the effects of varying 
parameters on gas mixing and disk destruction. 

\section{Parameter Study}

\subsection{Distance}

Various parameters were changed from the canonical case to study their effect on the survival of the disk and the injection 
efficiency of ejecta, including: the distance between the supernova and the disk, $d$; the explosion energy of 
the supernova, $E_{\rm ej}$; and the mass of gas in the disk, $M_{\rm disk}$.  In all these scenarios, the resolution stayed the same as in the canonical case.
The first parameter studied was the distance between the supernova and the disk.  
From the canonical distance of 0.3 pc, the disk was moved to 0.5 pc and 0.1 pc.  
The main effect of this change is to vary the density of the ejecta hitting the disk (see eq.\ [11]).
If the disk is closer, the gaseous ejecta is less diluted as it hits the disk.  
Hence these simulations are essentially equivalent to simulating a denser or a more tenuous clump of gas hitting the disk 
in an non-homogeneous supernova explosion.  
The results of these simulations can be seen in Table 2.  
The ``\% injected" column gives the percentage of the ejecta intercepted by the disk [with an assumed cross-section of $\pi$(30 AU)$^2$]  
that was actually mixed into the disk.  The third column gives the estimated $^{26}$Al/$^{27}$Al ratio that one would expect in  the 
disk if the SLRs were delivered {\bf in the gas phase}.  
This quantity was calculated using a disk chemical composition taken from Lodders (2003), and the ejecta isotopic composition from 
a 25 $\msol$ supernova taken from Woosley \&  Weaver (1995), which ejects $M=1.27 \times 10^{-4}\,\msol$ of ${}^{26}{\rm Al}$.   
Although the injection efficiency increases for denser ejecta, and the geometric dilution decreases for a closer supernova, 
gas-phase injection of ejecta into a disk at 0.1 pc cannot explain the SLR ratios in meteorites.  
The $^{26}$Al/$^{27}$Al ratio is off by roughly an order of magnitude from the measured value of 5 $\times$ 10$^{-5}$ 
(e.g., MacPherson et al.\ 1995).  
Stripping was more important with denser ejecta ($d = 0.1 \, {\rm pc}$), although still negligible compared to the mass 
of the disk;  only 0.7\% of the disk mass was lost.

\subsection{Explosion Energy}
  
We next varied the explosion energy, which defines the velocity at which the ejecta travel.  
The explosion energy was changed from 1 f.o.e.\ to 0.25 and 4 f.o.e., effectively modifying the ejecta velocity from 
2200 km/s to 1100 km/s and 4400 km/s, respectively.  
The results of the simulations can be seen in Table 3.  
Slower ejecta thermalizes to a lower temperature, and does not form such a strong reverse shock.  Therefore, slower 
ejecta is injected at a slightly higher efficiency into a disk.
Primarily, though, the results are insensitive to the velocity of the incoming supernova ejecta.

\subsection{Disk Mass}

The final parameter varied was the mass of the disk.  
From these simulations, the mass of the the minimum mass disk used in the canonical simulation was increased by a factor of 10,
and decreased by a factor of 10.   The results of the simulations can be seen in Table 4. 
Increasing the mass by a factor of 10 slightly increases, but this could be due to the fact that 
the disk does not get compressed as much as the canonical disk (it has a higher density and pressure at each radius).  Hence the disk has a larger surface to intercept the ejecta (the 
calculation for injection efficiency assumes a radius of 30 AU).  
Reducing the mass by a factor of 10 increases the efficiency.  
As the gas density in the disk is less, the pressure is less, and hence the ejecta is able to get closer to the midplane, 
increasing the amount injected.

\section {Conclusions}

In this paper, we have described a 2-D cylindrical hydrodynamics code we wrote, Perseus, and the results from the application
of this code to the problem of the interaction of supernova shocks with protoplanetary disks. 
A main conclusion of this paper is that disks are {\bf not} destroyed by a nearby supernova, even one as close as 0.1 pc. 
The robustness of the disks is a fundamentally new result that differs from previous 1-D analytical estimates (Chevalier 2000)
and numerical simulations (Ouellette et al.\ 2005). 
In those simulations, in which gas could not be deflected around the disk, the full momentum of the supernova ejecta was transferred
directly to each annulus of gas in the disk. 
Chevalier (2000) had estimated that disk annuli would be stripped away from the disk wherever 
$M_{\rm ej} V_{\rm ej} / 4\pi d^{2} > \Sigma_{\rm d} V_{\rm esc}$, where $\Sigma_{\rm d}$ is the surface density of the disk 
[$\Sigma_{\rm d} = 1700 \, (r / 1\, {\rm AU})^{-3/2} \, {\rm g} \, {\rm cm}^{-2}$ for a minimum mass disk; Hayashi et al.\ 1985), 
and $V_{\rm esc}$ is the escape velocity at the radius of the annulus. 
In the geometry considered here, the momentum is applied at right angles to the disk rotation, so $v_{\rm esc}$ can be 
replaced with the Keplerian orbital velocity, as the total kinetic energy would then be sufficient for escape. 
Also, integrating the momentum transfer over time (eq. [11]), we find $V_{\rm ej} = 3 v_{\star} / 4$.
Therefore, using the criterion of Chevalier (2000), and considering the parameters of the canonical case but with 
$d = 0.1 \, {\rm pc}$, the disk should have been destroyed everywhere outside of $30.2 \, {\rm AU}$, representing a loss 
of 13\% of the mass of a 40 AU radius disk.
Comparable conclusions were reached by Ouellette et al.\ (2005).

\begin{figure}[htbp]
\epsscale{1.0}
\begin{center}
\plotone{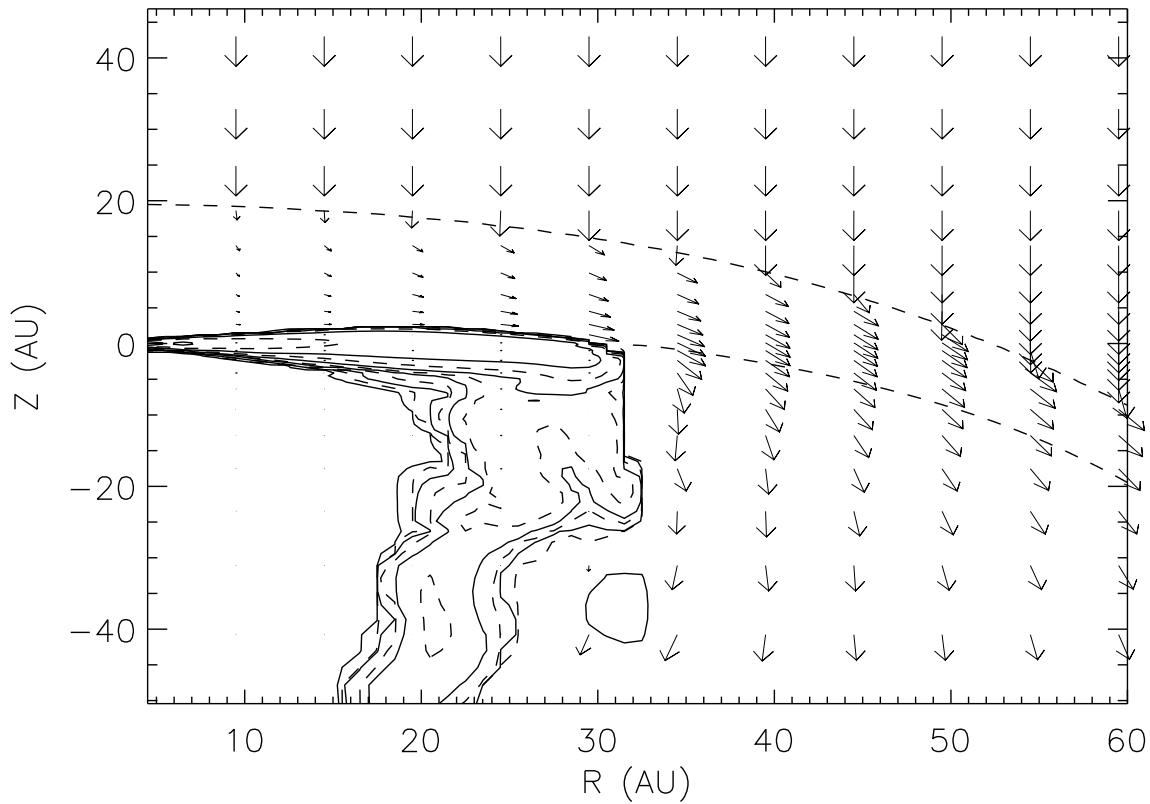}
\caption{Protoplanetary disk 50 years after first being hit by supernova ejecta.  The disk is substantially deformed
by the high pressures in the surrounding shocked gas.  The pressures compress the disk in the $z$ direction, and also
effectively aid gravity in the $r$ direction, allowing the gas to orbit at smaller radii with the same angular momentum.
Isodensity contours and velocity vectors as in Figure~\ref{Fig:5}.} 
\label{Fig:11}
\end{center}
\end{figure}

In contrast, as these 2-D simulations show, the disk becomes surrounded by high-pressure shocked gas that cushions the disk and
deflects ejecta around the disk.  
This high-pressure gas has many effects.  
First, the bow shock deviates the gas, making part of the ejecta that would have normally hit the disk flow around it.  
From Figure~\ref{Fig:11}, by following the velocity vectors, it is possible to estimate that the gas initially on trajectories
with$r > 20\,{\rm AU}$ will be deflected by $> 14^{\circ}$ after passing through the bow showk, and will miss the disk.  
For a disk 30 AU in size, this represents a reduction in the mass flux hitting by $\approx 45\%$; more thorough calculations 
give a reduction of $\approx 50\%$.  
Second, the bow shock reduces the forward velocity of the gas that does hit the disk. 
Gas deviated sideways about $14^{\circ}$, will have lost more than 10\% of its forward velocity upon reaching the disk.  
These two effects combined conspire to reduce the amount of momentum hitting the disk by 55\% overall.
By virtue of the smaller escape velocity and the lower disk surface density, gas at the disk edges is most vulnerable
to loss by the momentum of the shock, but it at the disk edges that the momentum of the supernova shock is most sharply
reduced.
Because of the loss of momentum, the disk in the previous paragraph could survive out to a radius of about $45 \, {\rm AU}$. 

A third, significant effect of the surrounding high-pressure shocked gas, though, is its ability to shrink the disk
to a smaller radius. 
The pressure in the post-shock gas is $\sim 2 \rho_{\rm ej}v_{\rm ej}^2/(\gamma +1) = 4.4 \times 10^{-4}\,{\rm dyne\, cm^{-2}}$,
so the average pressure gradient in the disk between about 30 and 35 AU is $\approx 1.9 \times 10^{-18} \,{\rm dyne\, cm^{-3}}$.  
This is to be compared to the gravitational force per volume at $35\,{\rm AU}$, 
$\rho g= 4.8 \times 10^{-19} \,{\rm dyne\, cm^{-3}}$ (at 35 AU, $\rho \sim 1.0 \times 10^{-15}$ in the canonical disk.)  
The pressure of the shocked gas enhances the inward gravitational force by a significant amount, causing gas of a given 
angular momentum to orbit at a smaller radius than it would if in pure Keplerian rotation. 
When this high pressure is relieved after the supernova shock has passed, the disk is restored to Keplerian rotation
and expands to its original size. 
While the shock is strongest, the high-pressure gas forces a protoplanetary disk to orbit at a reduced size, 
$\approx 30 \, {\rm AU}$, where it is invulnerable to being stripped by direct transfers of momentum. 
Because of these combined effects of the cushion of high-pressure shocked gas surrounding the disk---reduction in ejecta 
momentum and squeezing of the disk---protoplanetary disks even 0.1 pc from the supernova lose $< 1\%$ of their mass. 

\begin{figure}[htbp]
\epsscale{0.7}
\begin{center}
\plotone{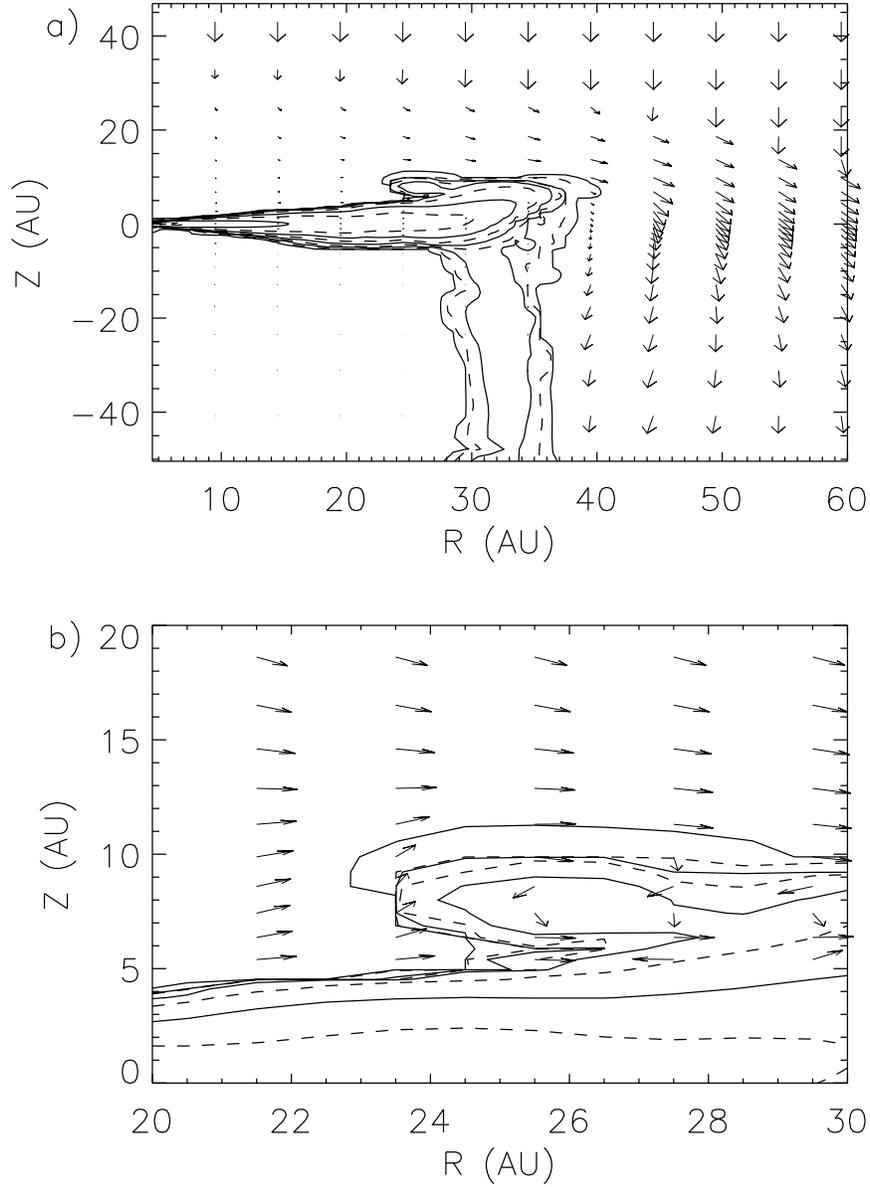}
\caption{(a) Protoplanetary disk 400 years after first being hit by supernova ejecta.  At this instant mass is being 
stripped off the top of the disk by a Kelvin-Helmholtz instability, seen in detail in (b).
Isodensity contours and velocity vectors as in Figure~\ref{Fig:5}.} 
\label{Fig:12} 
\end{center}
\end{figure}

\begin{figure}[htbp]
\epsscale{0.7}
\begin{center}
\plotone{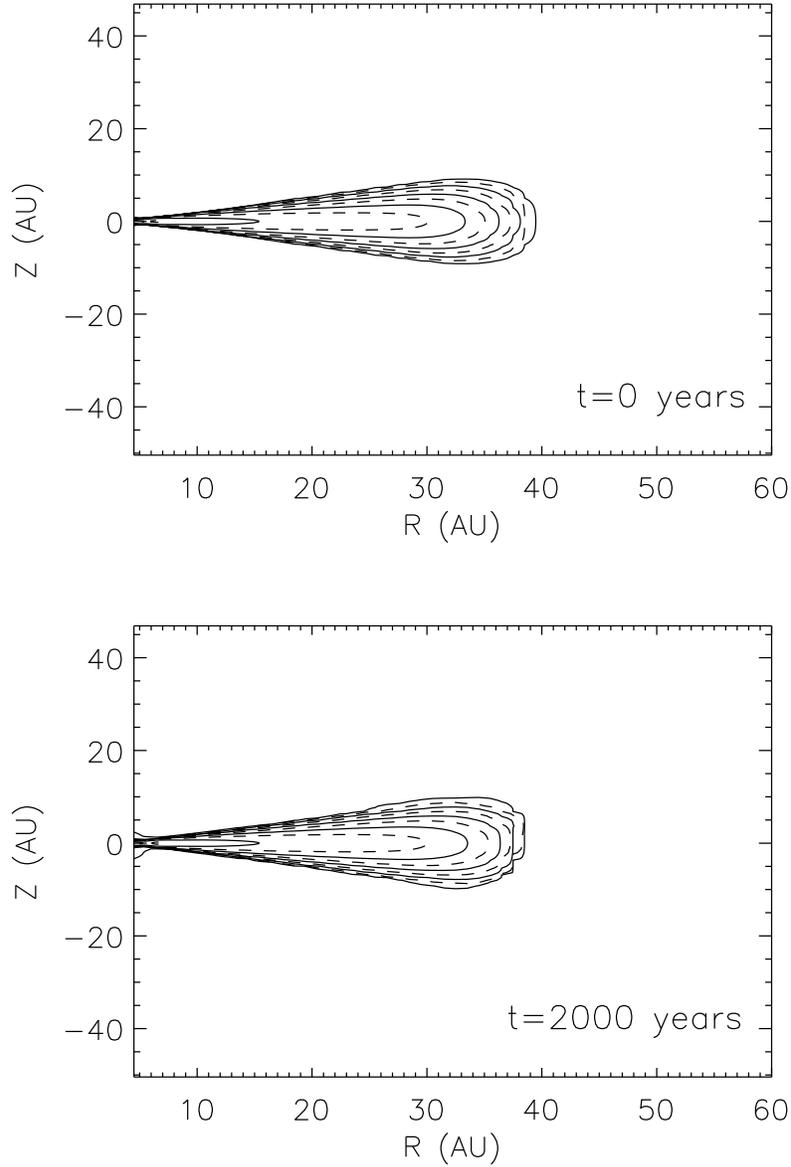}
\caption{(a) Protoplanetary disk prior to impact by supernova ejecta (same as the relaxed disk of Figure~\ref{Fig:4}),
and (b) 2000 years after first being struck by supernova ejecta.   This ``before and after" picture of the disk 
illustrates how the disk recovers almost completely from the shock of a nearby supernova.
Isodensity contours and velocity vectors as in Figure~\ref{Fig:5}.} 
\label{Fig:13}
\end{center}
\end{figure}

\begin{figure}[htbp]
\epsscale{0.8}
\begin{center}
\plotone{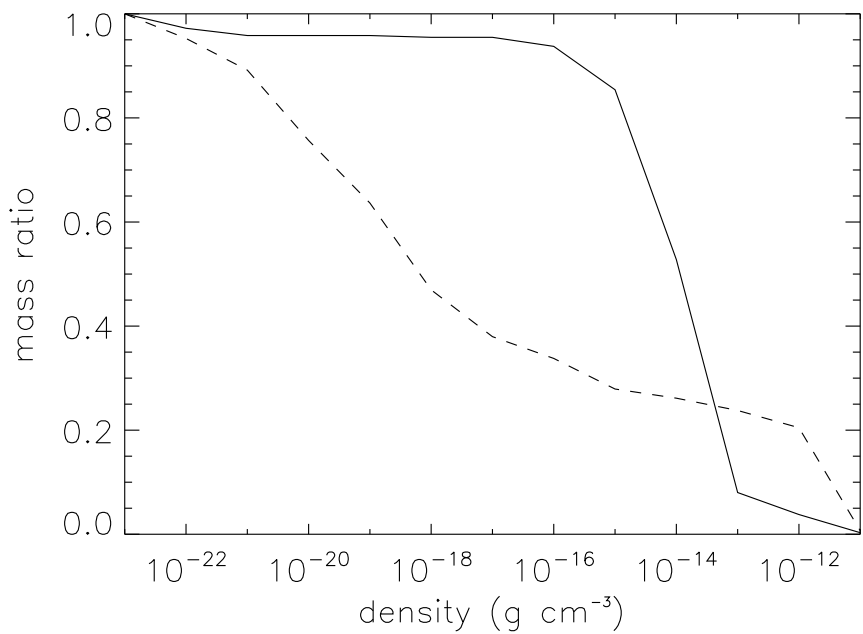}
\caption{Ratio of color mass to total mass within a given isodensity contour (abscissa).  The dashed line represents
the mass ratio after allowing the relaxed disk to evolve for 2000 years in the absence of a supernova; the solid line
represents the mass ratio after 2000 years of interaction with the supernova shock (our canonical simulation).
Supernova ejecta is injected very effectively up to densities where the shock would stall 
($\sim 10^{-14} \, {\rm g} \, {\rm cm}^{-3}$), much more effectively than can be accounted for by numerical 
diffusion alone.}
\label{Fig:14}
\end{center}
\end{figure}

\begin{figure}[htbp]
\epsscale{0.8}
\begin{center}
\plotone{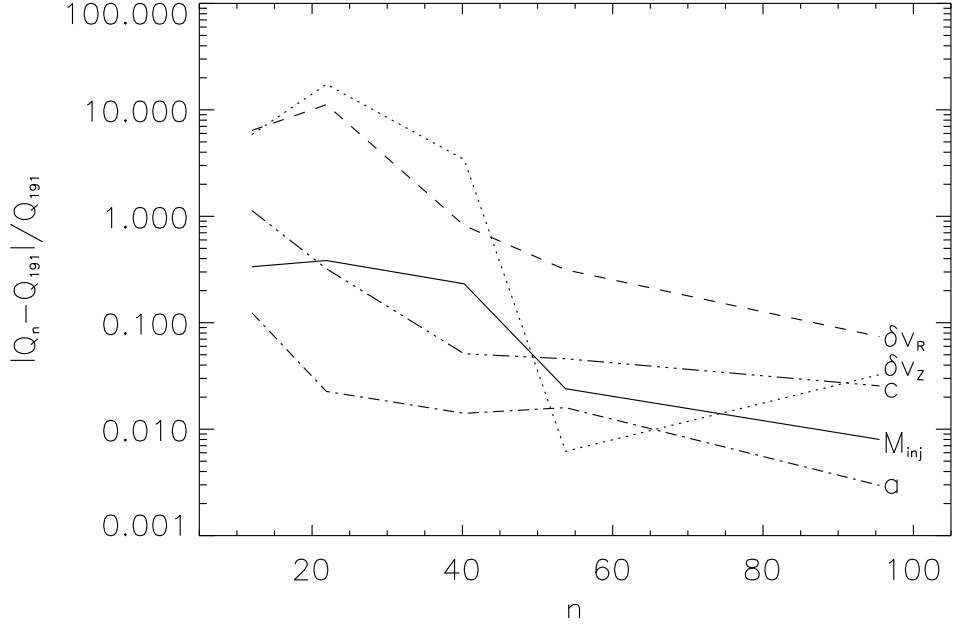}
\caption{Convergence properties of selected global variables.  The global variables are: the mass-weighted radius 
of the cloud, $a$; the mass-weighted cloud thickness, $c$; the dispersions in the mass-weighted radial ($\delta v_{r}$)
and vertical ($\vz$) velocities; and the mass of ejecta gas injected into the disk, $M_{\rm inj}$. 
The quantities are calculated at a time $t = 500$ years, but using 6 different numerical resolutions, 
$n = 12$, 22, 40, 54, 98 and 191. 
The deviation of each global quantity $Q$ from the highest-resolution value $Q_{191}$ is plotted against numerical 
resolution $n$.  
For our canonical simulation ($n = 98$), all quantities have converged at about the 10\% level.} 
\label{Fig:15} 
\end{center}
\end{figure}

\begin{figure}[htbp]
\epsscale{0.8}
\begin{center}
\plotone{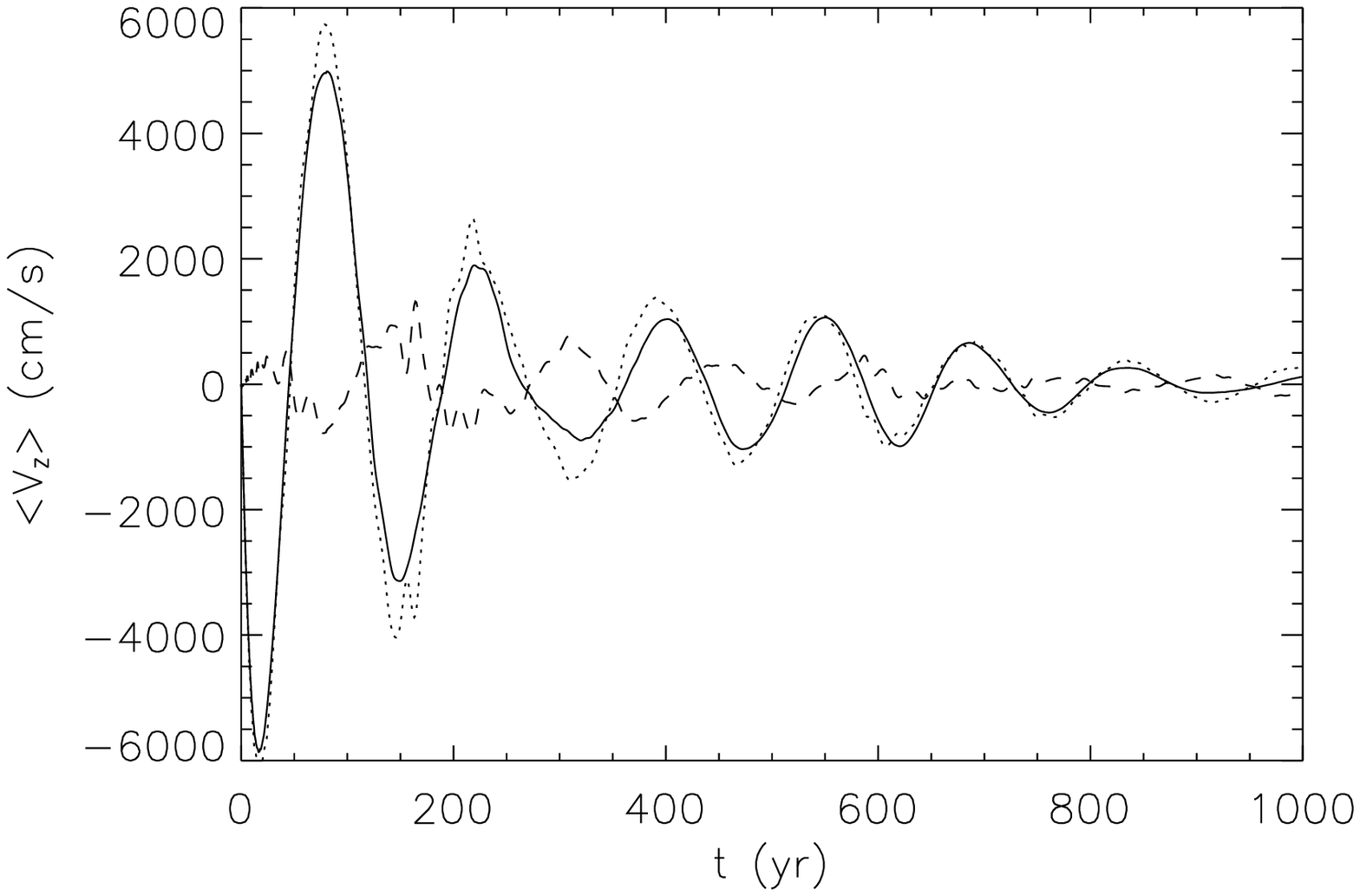}
\caption{Density-weighted velocity along the $z$ axis using the highest numerical simulation $n=191$ (solid line) and the
canonical resolution $n=98$ (dotted line).  The difference between them is plotted as the dashed line.  After absorbing 
the initial impulse of downward momentum from the supernova ejecta, the disk oscillates vertically about the position
of the central protostar with a period $\sim 150$ years, characteristic of the most affected gas at about 30 AU. }
\label{Fig:16} 
\end{center}
\end{figure}

Destruction of the disk, if it occurs at all, is due to stripping of the low-density upper layers of the disk
by Kelvin-Helmholtz (KH) instabilities.
We observed KH instabilities in all of our simulations (except $n=12$), and we observe their role in stripping gas from
the disk and mixing supernova ejecta into the disk (e.g., Figure~\ref{Fig:12}).
Our canonical numerical resolution ($n = 95$) and our highest-resolution simulation ($n = 191$), corresponding to
effectively 20-30 zones per cloud radius in the terminology of KMC, are just adequate to provide convergence at the 
10\% level, as described in \S 4.
We are confident we are capturing the relevant physics in our simulations, but we have shown that even if KH
instabilities are considerably more effective than we are modeling, that no more than about 1\% of the disk mass
could ever be affected by KH instabilities.
This is because the supernova shock stalls where the ram pressure is balanced by the pressure in the disk, and 
for typical protoplanetary disk conditions, this occurs several scale heights above the midplane.
It is unlikely that higher-resolution simulations would observe loss of more than $\sim 1\%$ of the disk mass.
We observed that the ratio of injected mass to disk mass was typically $\sim 1\%$ as well, for similar reasons.
We stipulate that mixing of ejecta into the disk is more subtle than stripping of disk mass, but given the limited
ability of the supernova ejecta to enter the disk, we find it doubtful that higher-resolution simulations would 
increase by more than a few the amount of gas-phase radionuclides injected into the disk.
Therefore, while disks like those observed in the Orion Nebula (McCaughrean \& O'Dell 1995) should survive the 
explosions of the massive stars in their vicinity, and while these disks would then contain some fraction of the 
supernova's gas-phase ejecta, they would not retain more than a small fraction ($\sim 1$\%) of the gaseous ejecta 
actually intercepted by the disk. 
If SLRs like ${}^{26}{\rm Al}$ are in the gas phase of the supernova (as modeled here), they will not be injected into the
disk in quantities large enough to explain the observed meteoritic ratios, failing by 1-2 orders of magnitude.

Of course, the SLRs inferred from meteorites, e.g., ${}^{60}{\rm Fe}$ and ${}^{26}{\rm Al}$, would not be detected
if they were not refractory elements.
These elements should condense out of the supernova ejecta as dust grains before colliding with a disk
(Ebel \& Grossman 2001).
Colgan et al.\ (1994) observed the production of dust at the temperature at which FeS condenses, 640 days after SN 1987A,
suggesting that the Fe and other refractory elements should condense out of the cooling supernova ejecta in less than a few 
years.
(The supernova ejecta is actually quite cool because of the adiabatic expansion of the gas.)
As the travel times from the supernova to the disks in our simulations are typically $20 - 500$ years, SLRs can be expected to
be sequestered into dust grains condensed from the supernova before striking a disk. 

Dust grains will be injected into the disk much more effectively than gas-phase ejecta. 
When the ejecta gas and dust, moving together at the same speed, encounter the bow shock, the gas is almost instantaneously
deflected around the disk, but the dust grains will continue forward by virtue of their momentum. 
The dust grains will be slowed only as fast as drag forces can act. 
The drag force $F$ on a highly supersonic particle is $F \approx \pi a^2 \, \rho_{\rm g} \, \Delta v^2$, where $a$ is the 
dust radius, 
$\rho_{\rm g}$ is the gas density, and $\Delta v$ is the velocity difference between the gas and the dust.
Assuming the dust grains are spherical with internal density $\rho_{\rm s}$, the resultant acceleration is 
$dv/dt = -(3\rho_{\rm g} \Delta v^2)/(4 \rho_{\rm s} a)$.  
Immediately after passage through the bow shock, the gas velocity has dropped to 1/4 of the ejecta velocity, so $\Delta v \approx (3/4) v_{\rm ej}$.  
Integrating the acceleration, we find the time $t_{1/2}$ for the dust to lose half its initial velocity:
\begin{equation}
t_{1/2} = \frac{16 \rho_{\rm s} a}{9 \rho_{\rm g} v_{\rm ej}}.
\end{equation}
Measurements of SiC grains with isotopic ratios indicative of formation in supernova ejecta reveal typical radii of 
$a \sim 0.5 \, \mu{\rm m}$ (Amari et al.\ 1994; Hoppe et al.\ 2000).
Assuming similar values for all supernova grains, and an internal density $\rho_{\rm s} = 2.5\, {\rm g\,cm^{-3}}$,
and using the maximum typical gas density in the region between the bow shock and the disk, 
$\rho_{\rm g} \approx 5 \times 10^{-20} \, {\rm g}\, {\rm cm}^{-3}$ , we find a minimum dust stopping time 
$t_{1/2} \approx 2 \times 10^{7} \, {\rm s}$.
In that time, the dust will have travelled about $300\,{\rm AU}$.
As the bow shock lies about 20 AU from the disk, the dust will encounter the protoplanetary disk well before 
travelling this distance, and we conclude that the dust the size of typical supernova SiC grains 
is not deflected around the disk.  
We estimate that nearly all the dust in the ejecta intercepted by the disk will be injected into the disk.
With nearly 100\% injection efficiency, the abundances of ${}^{26}{\rm Al}$ and ${}^{60}{\rm Fe}$ 
in a disk 0.15 pc from a supernova would be ${}^{26}{\rm Al}/{}^{27}{\rm Al} = 6.8 \times 10^{-5}$ and 
$\feratio = 4.8 \times 10^{-7}$ (using the yields from Woosley \&  Weaver 1995).
These values compare quite favorably to the meteoritic ratios (${}^{26}{\rm Al}/{}^{27}{\rm Al}=5.0 \times 10^{-5}$ and $\feratio = 3 - 7  \times 10^{-7}$; MacPherson et al.\ 1995, Tachibana \& Huss 2003), and we conclude that 
injection of SLRs into an already formed protoplanetary disk by a nearby supernova is a viable mechanism for delivering 
radionuclides to the early Solar System, provided the SLRs have condensed into dust.
In future work we will present numerical simulations of this process (Ouellette et al.\ 2007, in preparation).

\acknowledgments

We thank an anonymous referee for two very thorough reviews that significantly improved the manuscript. 
We also thank Chris Matzner for helpful discussions.

\newpage

{\small
\begin{table}[htbp]
Table 1 {\it Mass injected} \\
\begin{tabular}{cccc} \\
\# of zones ($r \times z$) & \% injected  \\ 
\hline 
8$\times$18 & 0.83 \\
16$\times$30 & 0.77 \\
30$\times$54 & 0.96 \\
39$\times$74 & 1.28 \\
60$\times$88 & 1.31 \\
\bf 76 $\times$120 & \bf 1.26 \\
152$\times$240 & 1.25 \\
\hline
\end{tabular}
\end{table}
}

\newpage

{
\small
\begin{table}[htbp]
Table 2 {\it Effect of Distance} \\
\begin{tabular}{cccc} \\
$d$ & \% injected & $^{26}$Al/$^{27}$Al\\ 
\hline 
0.1 pc & 4.3 & 6.4 $\times$ 10$^{-6}$  \\
\bf 0.3 pc & \bf 1.3 & \bf 2.2 $\times$ 10$^{-7}$\\
0.5 pc & 1.0 & 5.6 $\times$ 10$^{-8}$\\

\hline
\end{tabular}
\end{table}
}

\newpage

{
\small
\begin{table}[htbp]
Table 3 {\it Effect of Explosion Energy} \\
\begin{tabular}{cccc} \\
$E_{\rm ej}$ &  \% injected & $^{26}$Al/$^{27}$Al \\ 
\hline 
4.0  f.o.e.  & 1.0 & 1.7 $\times$ 10$^{-7}$\\
\bf 1.0  f.o.e.  & \bf 1.3 & \bf 2.2 $\times$ 10$^{-7}$\\
0.25 f.o.e.  & 1.7 & 2.8 $\times$ 10$^{-7}$\\

\hline
\end{tabular}
\end{table}
} 

\newpage

{
\small
\begin{table}[htbp]
Table 4 {\it Effect of Disk Mass} \\
\begin{tabular}{cccc} \\
$M_{\rm disk}$& \% injected & $^{26}$Al/$^{27}$Al\\ 
\hline 
0.084 $\msol$  & 1.4 & 2.3 $\times$ 10$^{-8}$\\
\bf 0.0084 $\msol$  & \bf 1.3 & \bf 2.2 $\times$ 10$^{-7}$\\
0.00084 $\msol$ & 2.2 & 3.6 $\times$ 10$^{-6}$\\

\hline
\end{tabular}
\end{table}
} 

\newpage 

\newpage 

\newpage 

\newpage 

\newpage 

\newpage 

\newpage 

\newpage 

\newpage 

\newpage 

\newpage 

\newpage 

\newpage 

\newpage

\newpage 

\newpage 

\begin{thebibliography}{}

\bibitem[]{1} Adams, F. C. \& Laughlin, G., 2001, Icarus, 150, 151
\bibitem[]{2} Amari, S., Lewis, R. S. \& Anders, E., 1994, GeCoA, 58, 459
\bibitem[Bedogni \& Woodward(1990)]{1990A&A...231..481B} Bedogni, R., \& Woodward, P.~R.\ 1990, \aap, 231, 481 
\bibitem[]{3} Cameron, A. G. W. \& Truran, J. W., 1977, Icarus, 30, 447
\bibitem[]{4} Chevalier, R. A., 2000, \apj, 538, 151
\bibitem[]{5} Clayton, D. D., 1977, Icarus, 32, 255
\bibitem[]{6} Colgan, S. W. J., Haas, M. R., Erickson, E. F., Lord, S. D. \& Hollenbach, D. J., 1994, \apj, 427, 874
\bibitem[D'Alessio et al.(2006)]{2006ApJ...638..314D} D'Alessio, P., Calvet, N., Hartmann, L., 
Franco-Hern{\'a}ndez, R., \& Serv{\'{\i}}n, H.\ 2006, \apj, 638, 314 
\bibitem[]{7} de Marco, O., O'Dell, C. R., Gelfond, P., Rubin, R. H. \& Glover, S. C. O., 2006, \aj, 131,2580 
\bibitem[Ebel \& Grossman(2001)]{2001GeCoA..65..469E} Ebel, D.~S., \& Grossman, L.\ 2001, \gca, 65, 469 
\bibitem[]{8} Goswami, J. N., Marhas, K. K., Chaussidon, M., Gounelle, M. \& Meyer, B. S., 2005, , Chondrites and the Protoplanetary Disk, ed. A. N. Krot, E. R. D. Scott and B. Reipurth (San Francisco: Astronomical Society of the Pacific), 485
\bibitem[]{9} Gounelle, M., Shu, F. H., Shang, H., Glassgold, A. E., Rehm, K. E.\& Lee, T., 2006, \apj, 640,1163
\bibitem[]{10} Harper, C. L., Jr., 1996, \apj, 466, 1026
\bibitem[]{11} Hayashi, C., Nakazawa, K. \& Nakagawa, Y., 1985, Protostars and Planets II , Tucson, AZ, University of Arizona Press, 1100
\bibitem[]{12} Hawley, J. F., Wilson, J. R. \& Smarr, L. L., 1984, \apj, 277, 296
\bibitem[]{13} Hester, J. J. \& Desch, S. J., 2005, Chondrites and the Protoplanetary Disk, ed. A. N. Krot, E. R. D. Scott and B. Reipurth (San Francisco: Astronomical Society of the Pacific), 107
\bibitem[]{14} Hoppe, P., Strebel, R., Eberhardt, P., Amari \& S., Lewis, R. S., 2000, M\&PS, 35, 1157
\bibitem[Hamuy et al.(1988)]{1988AJ.....95...63H} Hamuy, M., Suntzeff, N.~B., Gonzalez, R., \& Martin, G.\ 1988, \aj, 95, 63 
\bibitem[]{15} Huss, G. R. \& Tachibana, S., 2004, LPI, 35, 1811
\bibitem[]{16} Jacobsen, S. B., 2005, Chondrites and the Protoplanetary Disk, ed. A. N. Krot, E. R. D. Scott and B. Reipurth. (San Francisco: Astronomical Society of the Pacific), 548
\bibitem[]{17} Johnstone, D., Hollenbach, D. \& Bally, J.,1998, \apj, 499, 758
\bibitem[]{18} Kastner, J. H. \& Myers, P. C., 1994, \apj, 421, 605
\bibitem[]{19} Klein, R. I., McKee, C. F \& Colella, P., 1994, (KMC) \apj, 420, 213
\bibitem[]{20} Lada, C. J. \& Lada, E. A., 2003, ARA\&A, 41, 57L
\bibitem[]{21} Lee, T., Shu, F. H., Shang, H., Glassgold, A. E. \& Rehm, K. E., 1998, \apj, 506, 898
\bibitem[]{22} Leya, I., Halliday, A. N. \& Wieler, R., 2003, \apj, 594,605
\bibitem[]{23} Lodders, K., 2003, \apj, 591, 1220
\bibitem[Looney et al.(2006)]{2006ApJ...652.1755L} Looney, L.~W., Tobin, J.~J., \& Fields, B.~D.\ 2006, \apj, 652, 1755 
\bibitem[Mac Low et al.(1994)]{1994ApJ...433..757M} Mac Low, M.-M., McKee, 
C.~F., Klein, R.~I., Stone, J.~M., \& Norman, M.~L.\ 1994, \apj, 433, 757 
\bibitem[]{24} MacPherson, G. J., Davis, A. M. \& Zinner, E. K., 1995, Meteoritics, 30, 365
\bibitem[]{25} Matzner, C. D. \& McKee, C. F., 1999, \apj, 510, 379
\bibitem[]{26} McCaughrean, Mark J.; O'Dell, C. Robert, 1996, \aj, 111, 1977
\bibitem[]{27} Meyer, B. S. \& Clayton, D. D.,  2000, SSRv, 92, 133
\bibitem[]{28} Meyer, B. S.,  2005, Chondrites and the Protoplanetary Disk, ed. A. N. Krot, E. R. D. Scott and B. Reipurth. (San Francisco: Astronomical Society of the Pacific), 515
\bibitem[]{29} Mostefaoui, S., Lugmair, G. W. \& Hoppe, P., 2005, \apj, 625, 271
\bibitem[]{30} Mostefaoui, S., Lugmair, G. W., Hoppe, P. \& El Goresy, A., 2004, NewAR, 48, 155
\bibitem[Nakamura et al.(2006)]{2006ApJS..164..477N} Nakamura, F., McKee, C.~F., Klein, R.~I., \& Fisher, R.~T.\ 2006, 
\apjs, 164, 477 
\bibitem[Nittmann et al.(1982)]{1982MNRAS.201..833N} Nittmann, J., Falle, 
S.~A.~E.~G., \& Gaskell, P.~H.\ 1982, \mnras, 201, 833 
\bibitem[]{31} Oliveira, J. M., Jeffries, R. D., van Loon, J. Th., Littlefair, S. P. \& Naylor, T., 2005, MNRAS, 358, 21
\bibitem[Orlando et al.(2005)]{2005A&A...444..505O} Orlando, S., Peres, G., Reale, F., Bocchino, F., Rosner, R., 
Plewa, T., \& Siegel, A.\ 2005, \aap, 444, 505 
\bibitem[]{32} Ouellette, N., Desch, S. J., Hester, J. J. \& Leshin, L. A., 2005, Chondrites and the Protoplanetary Disk, ed. A. N. Krot, E. R. D. Scott and B. Reipurth. (San Francisco: Astronomical Society of the Pacific), 527
\bibitem[]{33} Quitt\'e, G., Latkoczy, C., Halliday, A. N., Sch\"onb\"achler, M. \& G\"unther, D., 2005, LPI, 36, 1827
\bibitem[]{34} Sod, G. A., 1978, J. Comput. Phys., 27, 1
\bibitem[]{35} Smith, N., Bally, J \& Morse, J. A., 2003, \apj, 587,105
\bibitem[]{36} Stone, J. M. \& Norman, M. L., 1992 \apjs, 80, 753
\bibitem[]{37} Sutherland, R. S. \& Dopita, M. A., 1993, \apjs, 88, 253
\bibitem[]{38} Tachibana, S. \& Huss, G. R.,  2003, \apj, 588, 41
\bibitem[]{39} Tachibana, S., Huss, G. R., Kita, N. T., Shimoda, G. \& Morishita, Y., 2006, \apj, 639, 87
\bibitem[]{40} Vanhala H. A. T. \& Boss A. P., 2000 \apj 538, 911
\bibitem[]{41} Vanhala H. A. T. \& Boss A. P., 2002 \apj 575, 1144
\bibitem[]{42} Wasserburg, G. J., Gallino, R. \& Busso, M.,1998, \apj, 500, 189
\bibitem[]{43} Woosley, S. E.,Heger, A. \& Weaver, T. A., 2002, RvMP, 74,1015
\bibitem[]{44} Woosley, S. E. \& Weaver, T., 1995. \apjs, 101, 181
\bibitem[Xu \& Stone(1995)]{1995ApJ...454..172X} Xu, J., \& Stone, J.~M.\ 1995, \apj, 454, 172 
\end{thebibliography}
\end{document}